\documentclass{article}
\usepackage{authblk}
\usepackage[utf8]{inputenc}
\usepackage{cite}
\usepackage{graphicx}
\usepackage{subcaption}
\usepackage{amsmath}
\usepackage{float}
\usepackage{enumerate}

\usepackage[a4paper,  margin=0.5in]{geometry}

\title{Bifurcation Analysis and Propagation Conditions of Free-Surface Waves in Incompressible Viscous Fluids in Finite Depth}
\author[1]{Arash Ghahraman}
\author[2]{Gyula Bene}

\affil[1]{Department of Theoretical Physics, Eötvös University, Pázmány Péter sétány 1/A, 1117 Budapest, Hungary}

\date{March 19, 2023}

\begin{document}

\maketitle

\abstract{Viscous linear surface waves are studied at arbitrary wavelength, layer thickness, viscosity and surface tension. We find that in shallow enough fluids no surface waves can propagate. This layer thickness is determined for some fluids, water, glycerin and mercury. Even in any thicker fluid layers, propagation of very short and very long waves is forbidden. When wave propagation is possible, only a single propagating
  mode exists for a given horizontal wave number. In contrast, there are two
  types of non-propagating modes. One kind of them exists at all wavelength and material parameters, and there are infinitely many such modes for a given wave number, distinguished by their decay rates. The other kind of non-propagating mode that is less attenuated may appear in zero, one or two specimens. We notice the presence of two length scales as material parameters, one related to viscosity and the other to surface tension. We consider possible modes for a given material on the parameter plane layer thickness versus wave number and discuss bifurcations among different mode types. Motion of surface particles and time evolution of surface elevation is also studied at various parameters in glycerin, and a great variety of behaviour is found, including counterclockwise surface particle motion and negative group velocity in wave propagation.
}

\section{Introduction}

Surface wave propagation is a widely-discussed subject in mathematics, fluid mechanics, hydro-geology, coastal engineering, etc. \cite{ref01_Meur2015derivation,ref02_antuono2013damping}. Historically, most works have focused on inviscid fluids. Indeed,  
in most situations in coastal engineering, the assumption of inviscid flow leads to very accurate results. Yet there are other physical scenarios where the viscosity needs to be taken into account \cite{ref03_hunt1964viscous,ref04_joseph2004dissipation,ref05_chen1998viscous}. Moreover, there are many situations in which both viscosity and surface tension should be considered \cite{ref02_antuono2013damping,ref06_armaroli2018viscous,ref07_sajjadi2017exact}. When surfactants are present for instance, the viscosity of the fluid can significantly impact the dynamics of the capillary waves \cite{ref11_shen2018capillary, ref26_spivak2002free}.  

Various equations have been proposed to model this propagation of surface waves. The goal is to find reduced models, particularly in size on simplified domains with as little fields as possible, should they be valid only in an asymptotic regime \cite{ref01_Meur2015derivation}.  To construct an accurate viscous model, however, one has to start with a linear study. 

Here should be first mentioned Boussinesq \cite{ref12_boussinesq1895lois} and Lamb \cite{ref43_lamb1924hydrodynamics}, who studied the effect of viscosity on free surface waves. They focused on linearized NS equations on deep-water and computed the dispersion relation. 
Longuet-Higgins \cite{ref15_longuet1969action,ref16_longuet1992theory} found Lamb's coefficient by using a boundary layer model and inspecting the equivalence of this model with the theory on weakly-damped waves by Ruvinsky and Freiman \cite{ref28_ruvinsky1985improvement}. Lighthill \cite{ref17_lighthill1978waves} used the deep-water inviscid solution to approximate the dissipative terms inside the kinetic energy equation for a viscous fluid. 
By assuming small amplitude gravity waves in a constant depth of water, Hunt studied the rate of attenuation of wave amplitude in shallow water \cite{ref03_hunt1964viscous}. Although these results are valid for an arbitrary depth, surface tension is not considered in his studies.
 In 1975, Kakutani and Matsuuchi \cite{ref18_kakutani1975effect} started from the NS equations and performed a clean boundary layer analysis. First, they made a linear analysis that gave the dispersion relation distinguishing various regimes.

In this study, we aim to investigate the effect of viscosity and surface tension in fluids based on the linearized Navier-Stokes equations to describe the propagation of surface waves, without any restriction on the parameters. We also identify the damping regimes in viscous fluids in the presence of surface tension, such as critical damping, and investigate the bifurcation points. Finally, we introduce an evolution equation to model the dynamics of an incompressible viscous fluid. 

The paper is structured in the following manner: Section 2 presents a derivation and discussion of the dispersion relation which is the foundation of our subsequent investigations. Section 3 examines various modes, while section 4 focuses on the study of layer thickness. Particle motion at the surface is explored in section 5, and the time evolution of surface elevations is investigated in section 6. Finally, the concluding section 7 provides a discussion of the results obtained in the study.

\section{Dispersion Analysis}
We consider linear surface waves on a viscous, incompressible fluid layer of
finite, constant depth $h$. 
The coordinates $x$ and $y$ are horizontal, $z$ is vertical. The origin lies at
the undisturbed fluid surface. Suppose that the flow corresponding to the surface
wave does not depend on $y$. We start with the Ansatz
\begin{eqnarray}
u(x,z,t)=f'(z) e^{i(kx-\omega t)}\;\label{e1}\\
w(x,z,t)=-ikf(z) e^{i(kx-\omega t)}\label{e2}
\end{eqnarray}
for the horizontal $u$ and vertical $w$ velocity components, the prime denotes derivative with respect to its argument. $k$ is a real, positive wave number, while $\omega$ is in general complex valued, its imaginary part describing the damping. Note that the Ansatz automatically satisfies the incompressibility condition ${\bf \nabla v} =0$,
since \(f\exp(i(kx-\omega t)\) specifies the stream function.
The linearized Navier-Stokes equation may be written as
\begin{eqnarray}
\frac{\partial {\bf v}}{\partial t}-\nu \triangle  {\bf v} ={\bf
  \nabla}\left(-\frac{p}{\rho}-g z\right)\;.\label{e4}
\end{eqnarray}
Since the right hand side is a full gradient, we have 
\begin{eqnarray}
\frac{\partial }{\partial z}\left(\frac{\partial { u}}{\partial t}-\nu
\triangle  { u} \right)
=\frac{\partial }{\partial x}\left(\frac{\partial { w}}{\partial t}-\nu
\triangle  { w} \right)\;.\label{e5}
\end{eqnarray}
Putting Ansatz (\ref{e1}), (\ref{e2}) into Eq.(\ref{e5}), we obtain
\begin{eqnarray}
f''''+\left(i\frac{\omega}{\nu}-2k^2\right)f''-\left(i\frac{\omega}{\nu}-k^2\right)k^2f=0\;.
\label{e6}
\end{eqnarray}
Eq.(\ref{e6}) has exponential solutions $f=\exp\left(\kappa z\right)$. For the
exponent $\kappa$ we get
\begin{eqnarray}
\kappa^4+\left(i\frac{\omega}{\nu}-2k^2\right)\kappa^2-\left(i\frac{\omega}{\nu}-k^2\right)k^2f=0\;.
\label{e7}
\end{eqnarray}
The solutions are 
\begin{eqnarray}
\kappa_{1,2}=\pm k\;,\\
\kappa_{3,4}=\pm \sqrt{k^2-i\frac{\omega}{\nu}}\;.
\label{e8}
\end{eqnarray}
For brevity, we shall use the notation $\kappa$ for $\kappa_3=-\kappa_4$ and
$k$ for $\kappa_1=-\kappa_2$. The general solution for $f$ may be given as
\begin{eqnarray}
f=a_1\cosh\left[k(z+h)\right]+a_2\sinh\left[k(z+h)\right]+b_1\cosh\left[\kappa(z+h)\right]+b_2\sinh\left[\kappa(z+h)\right]\;,
\label{e9}
\end{eqnarray}
where $a_1$, $a_2$, $b_1$, $b_2$ are integration constants and $h$ stands for
the fluid depth. Then boundary conditions at the bottom,
\begin{eqnarray}
v_x(z=-h)=v_z(z=-h)=0\;,
\label{e10}
\end{eqnarray}
imply
\begin{eqnarray}
a_1+b_1=0\;,\label{e11}\\
a_2 k+b_2 \kappa=0\;,\label{e12}
\end{eqnarray}
or
\begin{eqnarray}
a_1=A\;,\quad b_2=B\;,\quad b_1=-A\;,\quad a_2=-\frac{\kappa}{k}B\;,
\label{e13}
\end{eqnarray}
expressed in terms of the new constants $A$ and $B$. Hence for $f$ we get
\begin{eqnarray}
f=A\cosh\left[k(z+h)\right]-\frac{\kappa}{k}B\sinh\left[k(z+h)\right]-A\cosh\left[\kappa(z+h)\right]+B\sinh\left[\kappa(z+h)\right]\;.
\label{e14}
\end{eqnarray}
Upon integrating the $x$ component of the Navier-Stokes equation  with respect
to $x$, we get the pressure as
\begin{eqnarray}
p=p_o-\rho g z-\rho  e^{i (kx-\omega t)}\left(-\frac{\omega}{k}f'-i\nu
k f'+i\frac{\nu}{k}f'''\right)\;.
\label{e15}
\end{eqnarray}
At the fluid surface we have the boundary conditions that the strain forces
are continuous, therefore (in linear approximation) we have 
\begin{eqnarray}
\frac{\partial v_x}{\partial z}+\frac{\partial v_z}{\partial x}=0
\label{e16}
\end{eqnarray}
for the shear and
\begin{eqnarray}
p-2\rho \nu \frac{\partial v_z}{\partial z}=p_0-\sigma
\frac{\partial^2\eta}{\partial x^2}\;.
\label{e17}
\end{eqnarray}
for the pressure. Here $\eta=\eta(x,t)$ stands for the deviation of the fluid
surface from equilibrium and $\sigma$ is the surface tension. 

Eq.(\ref{e16}) implies 
\begin{eqnarray}
f''+k^2f=0
\label{e18}
\end{eqnarray}
at $z=0$, while  Eq.(\ref{e17}) implies
\begin{eqnarray}
-g\eta + \frac{\sigma}{\rho}\frac{\partial^2\eta}{\partial x^2}- e^{i(kx-\omega t)}\left(-\frac{\omega}{k}f'-i\nu
k f'+i\frac{\nu}{k}f'''\right)+2i\nu k f' e^{i(kx-\omega t)} =0\;.
\label{e19}
\end{eqnarray}
We have at the surface (again in linear approximation)
\begin{eqnarray}
\frac{\partial \eta}{\partial t}=v_z\;.
\label{e20}
\end{eqnarray}
Note that on the right hand side we may set $z=0$. Putting here the expression
of $v_z$ (i.e., Eq.(\ref{e2})) and combining the result with
Eq.(\ref{e19}), we have 
\begin{eqnarray}
\left(1+\frac{\sigma}{g\rho}k^2\right)k^2f-\frac{1}{g}\left(\omega^2+3i\omega \nu k^2\right) f'+i\frac{\omega\nu}{g}f'''=0\;,
\label{e21}
\end{eqnarray}
or, in terms of $\kappa$ (cf. Eq.(\ref{e8}))
\begin{eqnarray}
\left(1+\frac{\sigma}{g\rho}k^2\right)k^2f+\frac{\nu^2}{g}(\kappa^2-k^2)(\kappa^2+2k^2)
f'-\frac{\nu^2}{g}(\kappa^2-k^2) f'''=0\;.
\label{e22}
\end{eqnarray}
Here again $z=0$.
Inserting now the solution (\ref{e14}) into Eqs.(\ref{e18}) and (\ref{e22}) we
obtain a linear homogeneous system of equation for the coefficients $A$ and
$B$. The vanishing of the determinant of this system (which is the condition
for the existence of a nontrivial solution) may be expressed in terms of the
dimensionless variables
\begin{eqnarray}
K=kh\label{e23a}\\
Q=\kappa h\label{e23b}\\
p=\frac{\nu^2}{gh^3}
\label{e23c}\\
s=\frac{\sigma}{\rho gh^2}
\label{e23d}
\end{eqnarray}
as
\begin{eqnarray}
&K\left( Q\sinh K \cosh Q-K\cosh K\sinh
Q \right)(1+sK^2)+p\left[ -4K^2Q\left(K^2+Q^2\right)\right.\nonumber \\&\left.+Q\left(Q^4+2K^2Q^2+5K^4\right) \cosh
  K\cosh Q -K\left( Q^4+6K^2Q^2+K^4\right) \sinh K\sinh Q \right]=0\;.
\label{e25}
\end{eqnarray}
Note that $1/\sqrt{p}=\sqrt{gh}h/\nu$ looks like a Reynolds number, however, it has a quite different physical meaning, since it refers to the linear case.

Parameters $p$ and $s$ may be expressed in terms of the viscous
length scale 
\begin{eqnarray}
\ell_\nu = \left(\frac{ \nu^2}{g}\right)^{1/3}
\label{e25a}
\end{eqnarray}
and the length scale related to surface tension
\begin{eqnarray}
\ell_\sigma = \left(\frac{ \sigma}{\rho g}\right)^{1/2}\;,
\label{e25b}
\end{eqnarray}
namely,
\begin{eqnarray}
p=\left(\frac{\ell_\nu}{h}\right)^3
\label{e25c}
\end{eqnarray}
and
\begin{eqnarray}
s=\left(\frac{\ell_\sigma}{h}\right)^2\;.
\label{e25d}
\end{eqnarray}
It is obvious that the ratio
\begin{eqnarray}
  \mu=\frac{\ell_\sigma}{\ell_\nu}=\frac{s^{1/2}}{p^{1/3}}
\label{e25e}
\end{eqnarray}
is a material parameter (apart from \(g\)), and does not depend on the layer
thickness. On the other hand, when parameter dependence is studied, often this
is done by changing the layer thickness of a given material. In that case it
is advisable to use parameters \(\mu\) and \(p\), while
\begin{eqnarray}
  s=\mu^2 p^{2/3}\;.
\label{e25f}
\end{eqnarray}

Given parameters $p$ and $s$ (or \(p\) and \(\mu\)), and scaled wave number $K$, a solution $Q$ of
Eq.(\ref{e25}) yields the angular frequency (cf. Eqs.(\ref{e8}), (\ref{e23b}) )
\begin{eqnarray}
\omega=-i\frac{\nu}{h^2} \left(K^2-Q^2\right) \;.
\label{e26}
\end{eqnarray}
For the ratio of the coefficients $A$ and $B$ we get (cf. Eqs.(\ref{e14}), (\ref{e18})) 
\begin{eqnarray}
\frac{B}{A}=\frac{2K^2\cosh(K)-(K^2+Q^2)\cosh(Q)}{2Q K\sinh(K)-(K^2+Q^2)\sinh(Q)}\;.
\label{e26a}
\end{eqnarray}

Note that if $Q$ is a solution of Eq.(\ref{e25}), so is $-Q$. On the other
hand, this sign does not matter when calculating $\omega$ or $f$ (cf. Eq.(\ref{e14})). Henceforth we
assume that the real part of $Q$ is positive, and thus $\tanh Q \rightarrow
1$ when $|Q|\rightarrow \infty$.
\subsection{Small viscosity case}
In the small viscosity case Eq.(\ref{e25}) may be solved approximately. In
that case $p\rightarrow 0$ and $|Q|\rightarrow \infty$. This implies that in
leading order Eq.(\ref{e25}) reduces to
\begin{eqnarray}
K\tanh K\;(1+sK^2) +p Q_0^4=0\;,
\label{e27}
\end{eqnarray}
or 
\begin{eqnarray}
Q_0^2=- i\sqrt{\frac{K\tanh K\;(1+sK^2)}{p}}\;.
\label{e28}
\end{eqnarray}
Here the negative sign has been chosen in order to get positive real part of angular
frequency via Eq.(\ref{e26}). Further, according to the convention mentioned above,
we have
\begin{eqnarray}
Q_0=\frac{1-i}{\sqrt{2}}\left(\frac{K\tanh K\;(1+sK^2)}{p}\right)^{1/4}\;.
\label{e29}
\end{eqnarray}
A systematic expansion in terms of $p^{1/4}$ leads in the next two orders to

\begin{eqnarray}
 Q=Q_0-\frac{K}{2\sinh(2K)}-\frac{K^2}{2Q_0}\frac{Y^2+6Y+5}{Y(Y+4)} \;.
\label{e31}
\end{eqnarray}
Here $Y=4\sinh^2K$.
To this order we have for the angular frequency
\begin{eqnarray}
\omega&=&\left[\sqrt{\left(gk+\frac{\sigma k^3}{\rho}\right)\tanh (kh)}-\frac{\sqrt{2\nu k^2\sqrt{\left(gk+\frac{\sigma k^3}{\rho}\right)\tanh (kh)}}}{2\sinh(2kh)}\right]\nonumber\\&-i&\left[\frac{\sqrt{2\nu k^2\sqrt{\left(gk+\frac{\sigma k^3}{\rho}\right)\tanh (kh)}}}{2\sinh(2kh)}+2\nu k^2\frac{Y^2+5Y+2}{Y(Y+4)}\right]\;.
\label{e32}
\end{eqnarray}
The first term of the real part
is the well known dispersion relation of surface waves in ideal fluids. As for
damping, the leading term is the first one in the second bracket, proportional
to $\sqrt{\nu}$, except in
deep fluid. In deep fluid ($kh\rightarrow \infty$) this term vanishes and one
gets the well know damping exponent $2\nu k^2$. The result (\ref{e32}) have first been
published (for $\sigma=0$) in \cite{ref45_biesel1949calculation} and then to higher orders in
\cite{ref03_hunt1964viscous}. Note that taking into account surface tension is formally
equivalent with replacing $p$ with $p/(1+sK^2)$.
\subsection{Behaviour at large wavelengths}
At \(K=0\) Eq.(\ref{e25}) reduces to
\begin{eqnarray}
    Q^5\cosh(Q)=0
\end{eqnarray}
(independently of \(p\) and \(s\)), that can be solved analitically:
either
\begin{eqnarray}
    Q=0\label{xq1}
\end{eqnarray}
or
\begin{eqnarray}
    Q=i(2n+1)\frac{\pi}{2}\;,\;n=0,1,2,\dots\label{xq2}
\end{eqnarray}
The corresponding frequencies are \(\omega=0\) and
\begin{eqnarray}
\omega=-i\frac{\nu\pi^2}{4h^2} (2n+1)^2 \;.
\label{xq26}
\end{eqnarray}
Note that the order of the limits \(\nu\rightarrow 0\) and \(K\rightarrow 0\) does matter. If we take the limit \(\nu\rightarrow 0\) first, we get ideal fluid, while taking  \(K\rightarrow 0\) first, we get a limit where viscosity dominates and no wave propagation is possible. 

Eqs.(\ref{xq1}), (\ref{xq2}) proved to be important technically, as solutions  Eq.(\ref{e25}) could be obtained numerically from the differential equation

\begin{equation}
    \frac{dQ}{dK} = -\frac{\partial D/ \partial K}{\partial D/ \partial Q} \label{eq45}
\end{equation}
while Eqs.(\ref{xq1}), (\ref{xq2}) play the role of initial conditions at K=0. At bifurcations we get singular behaviour, that is avoided on the complex \(K\) plane, afterwards we return to real \(K\) values. This method worked, one even could choose different branches by going round the singularity from left of right on a half circle, yet it was somewhat in a state-of-art, as we had to experiment to get the correct radius of half-circles. Another method we applied to get Q was direct numerical solution of  
Eq.(\ref{e25}). In this case bracketing solutions was a nontrivial task. Hence, visualizing roots was helpful.   We plotted those lines in the complex Q-plane, where the real part (in blue) or the imaginary part (in red)  of determinant \(D\) vanishes. The crossing of these lines correspond to a solution for \(Q\) (see Fig.\ref{qmod1}). Note that we compared the results obtained with these two different methods, and found an excellent agreement.

\subsection{Numerical results for frequencies}
\begin{figure}[H]
\centering
\includegraphics[width=12.5 cm]{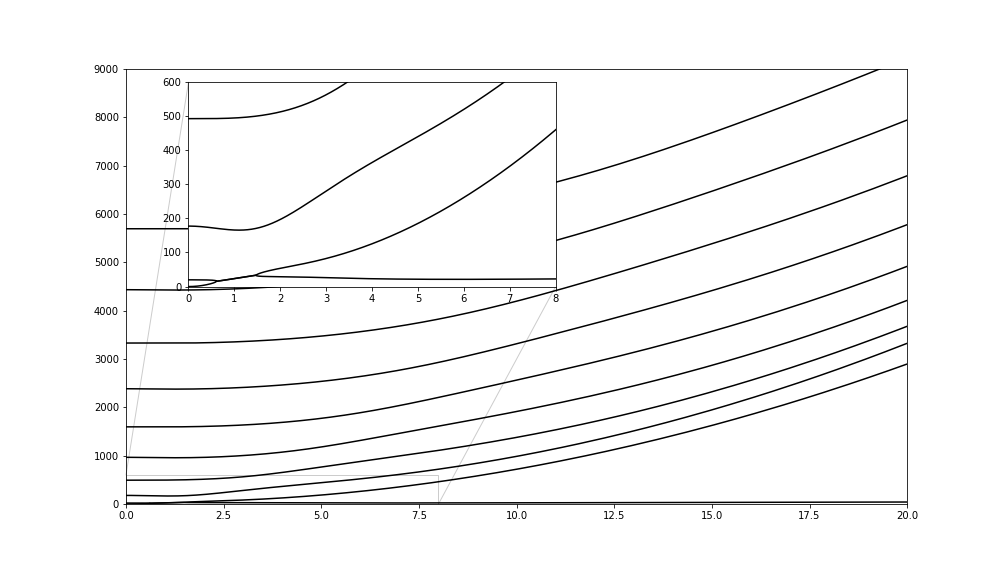}
\caption{First 10 branches of solutions for glycerin at $p=0.077$. The majority of the frequencies remain purely imaginary, but the two lowest branches intersect when the value of K is increased.\label{figx1}}
\end{figure}

\begin{figure}[H]
     \centering
     \begin{subfigure}{\textwidth}
         \centering
         \includegraphics[width=10cm]{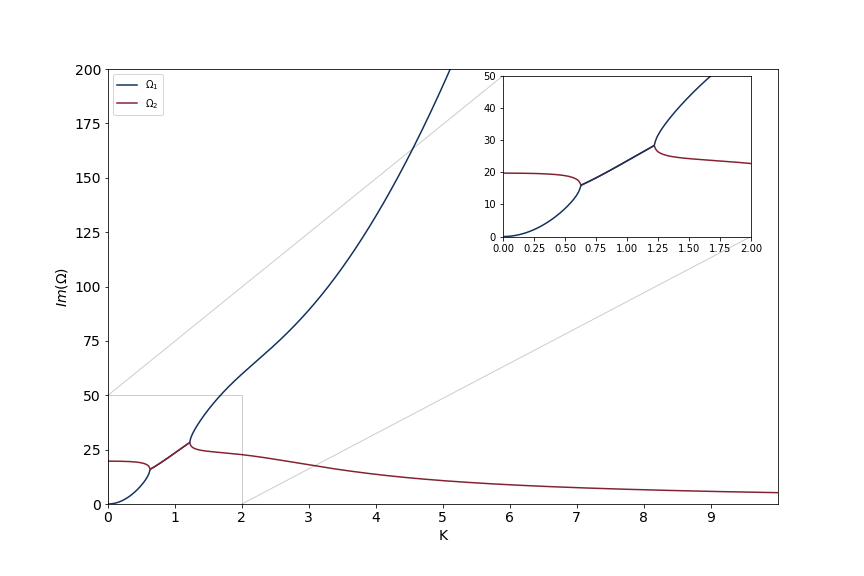}
         \caption{Two lowest branches of solutions for glycerin at $p=0.077$.}
          \label{fig2a}
     \end{subfigure}

     \begin{subfigure}{\textwidth}
         \centering
         \includegraphics[width=10cm]{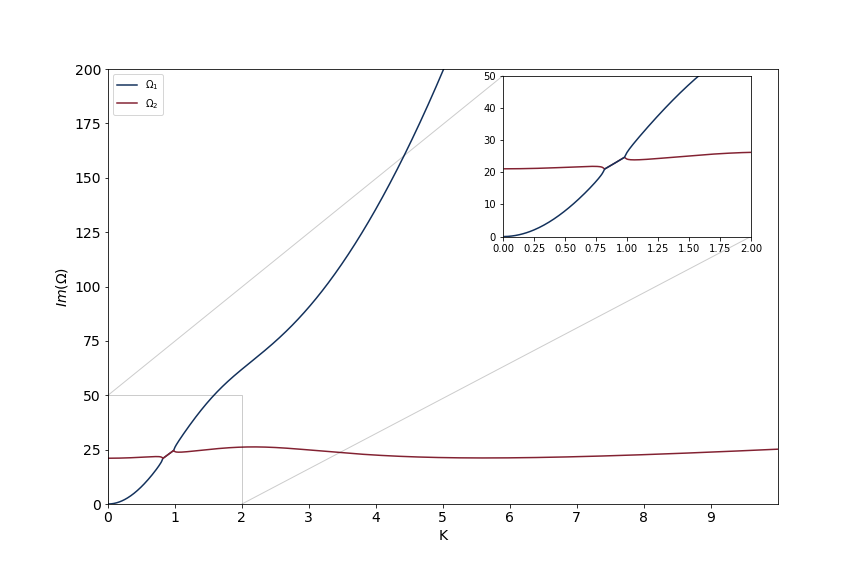}
         \caption{Two lowest branches of solutions for glycerin at $p=0.085$.}
            \label{fig2b}
     \end{subfigure}

     \begin{subfigure}{\textwidth}
         \centering
         \includegraphics[width=10cm]{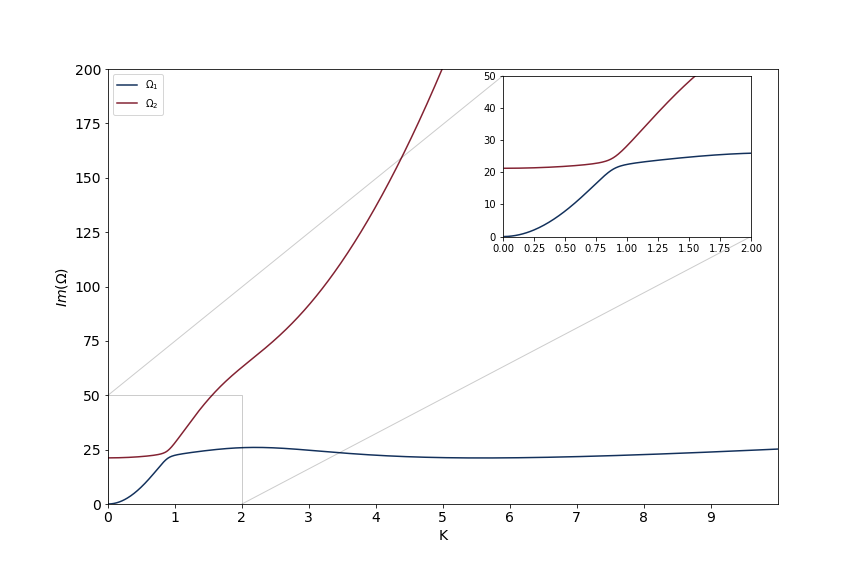}
         \caption{Two lowest branches of solutions for glycerin at $p=0.086$. The collision no longer occurs by increasing the value of p. }
            \label{fig2c}
     \end{subfigure}
     
\caption{Two lowest branches of solutions for glycerin at different p values. Moduli of imaginary parts of frequencies are displayed.}
\label{fig2}
\end{figure}

\begin{figure}[H]
\centering
\includegraphics[width=12.5 cm]{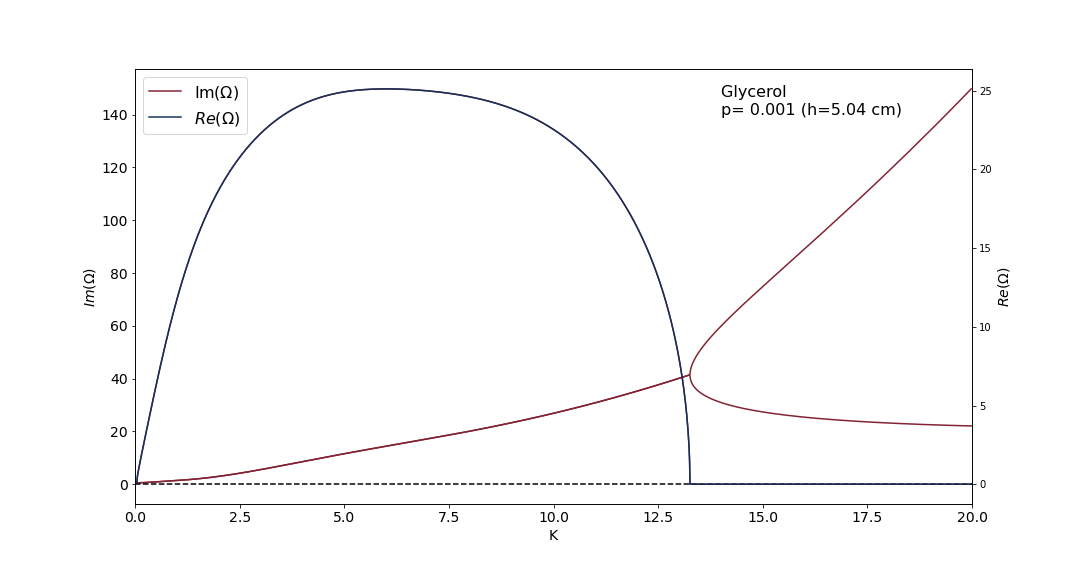}
\caption{The real and imaginary parts of frequencies corresponding to the lowest lying branches versus $K$ for glycerin at $p=0.001 (h=5.04 cm)$.}\label{fig_reim}
\end{figure}

Having obtained \(Q\) for parameters \(p\), \(s\) and \(K\) one may calculate mode frequencies. We present such results for a given material - glycerine -, at some selected \(p\) values and as a function of scaled wave number \(K\). The grand picture is this: even for fixed parameters one always obtains infinitely many solutions, organized into branches as \(K\) changes (cf. Fig.\ref{figx1}). They emanate from values (\ref{xq26}). Most of them remains purely imaginary, but the lowest two may collide when increasing \(K\). At such a collision a bifurcation takes place, like in  Figs.\ref{figx1},\ref{fig2a},\ref{fig_reim}, and two imaginary solutions may combine to complex solutions with nonzero real parts (cf. Fig.\ref{fig_reim}). In that case the imaginary parts are the same, and the real parts differ in their signs only.
As the value of \(p\) is increased, after a while such collision no longer occurs, like in Fig.\ref{fig2c}. The intermediate situation is approximately shown in Fig.\ref{fig2b}. 
The mechanism of this possible bifurcation will be studied in more details in the next section.

\section{Parameter dependence}

\subsection{Wave modes}
Plotting the real and imaginary parts of Eq.(\ref{e25}) on the complex $Q$
plane one usually observes several intersections, i.e.,
roots. They can be real, purely imaginary, or complex with nonzero real and
imaginary parts. In the first two cases the angular frequency (\ref{e26}) is
purely imaginary, so these modes decay exponentially with time. Propagating
modes are only possible if $Q$ is complex.
\subsubsection{Modes with real Q}
In this case Eq.(\ref{e26}) implies that $Q<K$, since the decay rate cannot be negative. 
We have found numerically that at a given parameter settings ($s$, $p$ and $K$)
there can exist zero, one or two real modes.\footnote{Due to the symmetry of
the solutions, we consider roots only in the first quadrant.} 
There is always a trivial solution $Q=K$. With this, however, we get from
Eq.(\ref{e14}) $f=0$, so this is solution irrelevant.

In case of nontrivial real solutions the decay rate is always smaller
than $\nu k^2$. Velocity components may be calculated from Eqs.(\ref{e1}),
(\ref{e2}), (\ref{e14}) and (\ref{e26a}). Provided that coefficient $A$ is
real, $v_x$ is real, too, while $v_z$ is purely imaginary, so there is a
$90^\circ$ phase shift in their $x$-dependent oscillations.

 \subsubsection{Modes with imaginary Q}
In this case the decay rate is always larger
than $\nu k^2$. Such modes exist at any parameter setting, moreover, there are infinitely
many of them. Indeed, if $Q$ is purely imaginary and its modulus is large,
Eq.(\ref{e25}) reduces to 
\begin{eqnarray}
p\left[ Q^5 \cosh
  K\cosh Q -K Q^4\sinh K\sinh Q \right]=0\;,
\label{f25}
\end{eqnarray}
or, substituting $Q=iw$,
\begin{eqnarray}
w \cosh
  K\cos w -K \sinh K\sin w =0\;.
\label{f25a}
\end{eqnarray}
Now, if $w=2n\pi$ ($n$ being an integer), the left hand side is positive, and
if $w=2n\pi+\pi/2$, the left hand side is negative. Therefore, between these
values there is a root for any (arbitrarily large) $n$. The distance between
imaginary roots is approximately constant and independent of viscosity.

As before, the phase of velocity
components does not change with depth, while there is a $90^\circ$ phase shift
between the $x$ and $z$ components. It is interesting that imaginary $Q$
causes an oscillatory behavior with depth.

\subsubsection{Modes with complex Q}
On the basis of our numerical investigations, we believe that
 at a given parameter setting at most one such mode can exist. If there
 exists one, then at the same parameter setting no mode with real $Q$ can exist. 
This time the phases of velocity
components do change with depth. 
An oscillatory dependence on depth is in principle present, but much less
pronounced than in the imaginary case.

\begin{figure}[!ht]
\centering
\subfloat[$s$=7.44, $p$=2.0 and $K$=2.0\label{m:c}]{
  \includegraphics[width=45mm]{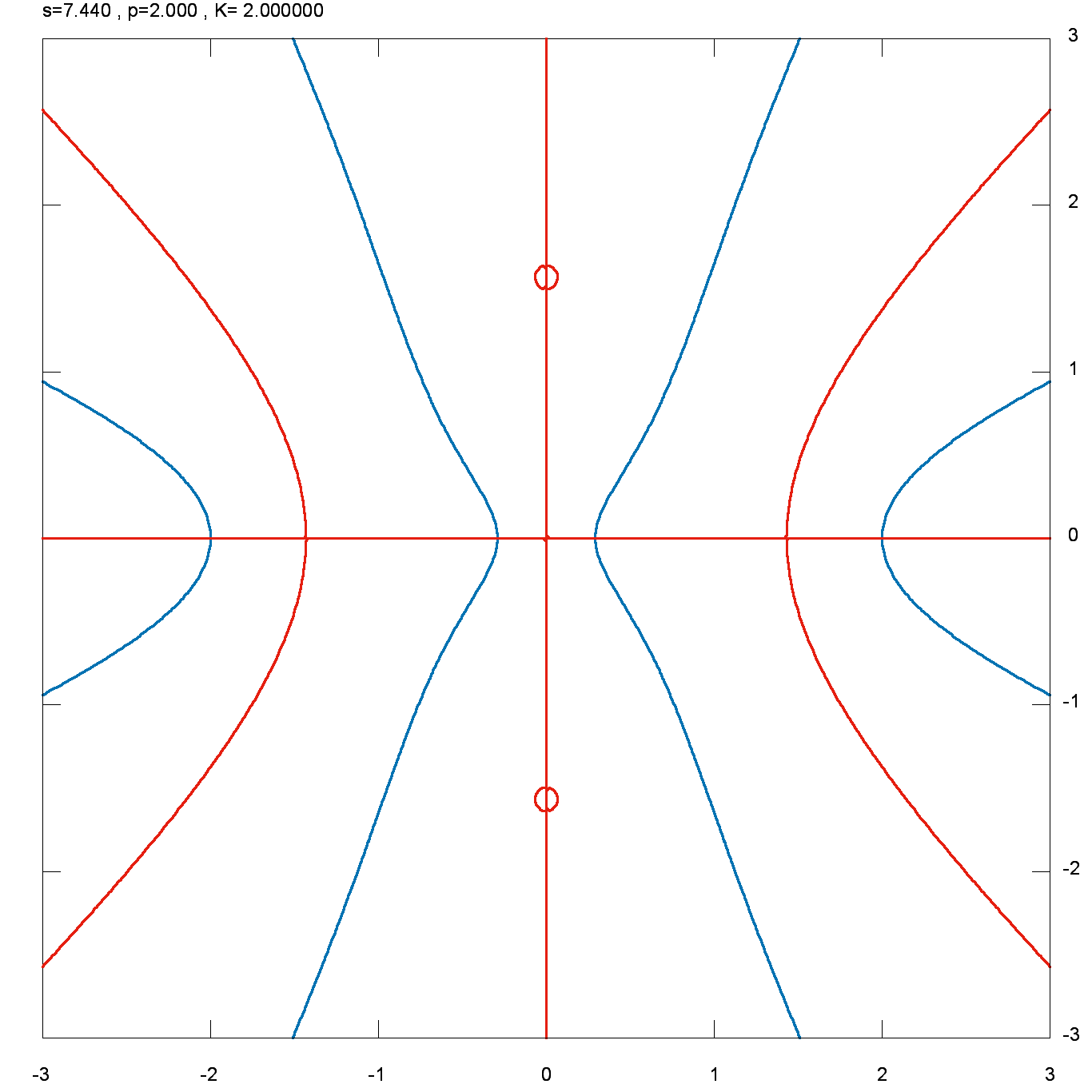}
}
\subfloat[$s$=7.44, $p$=2.0 and $K$=2.021\label{m:d}]{
  \includegraphics[width=45mm]{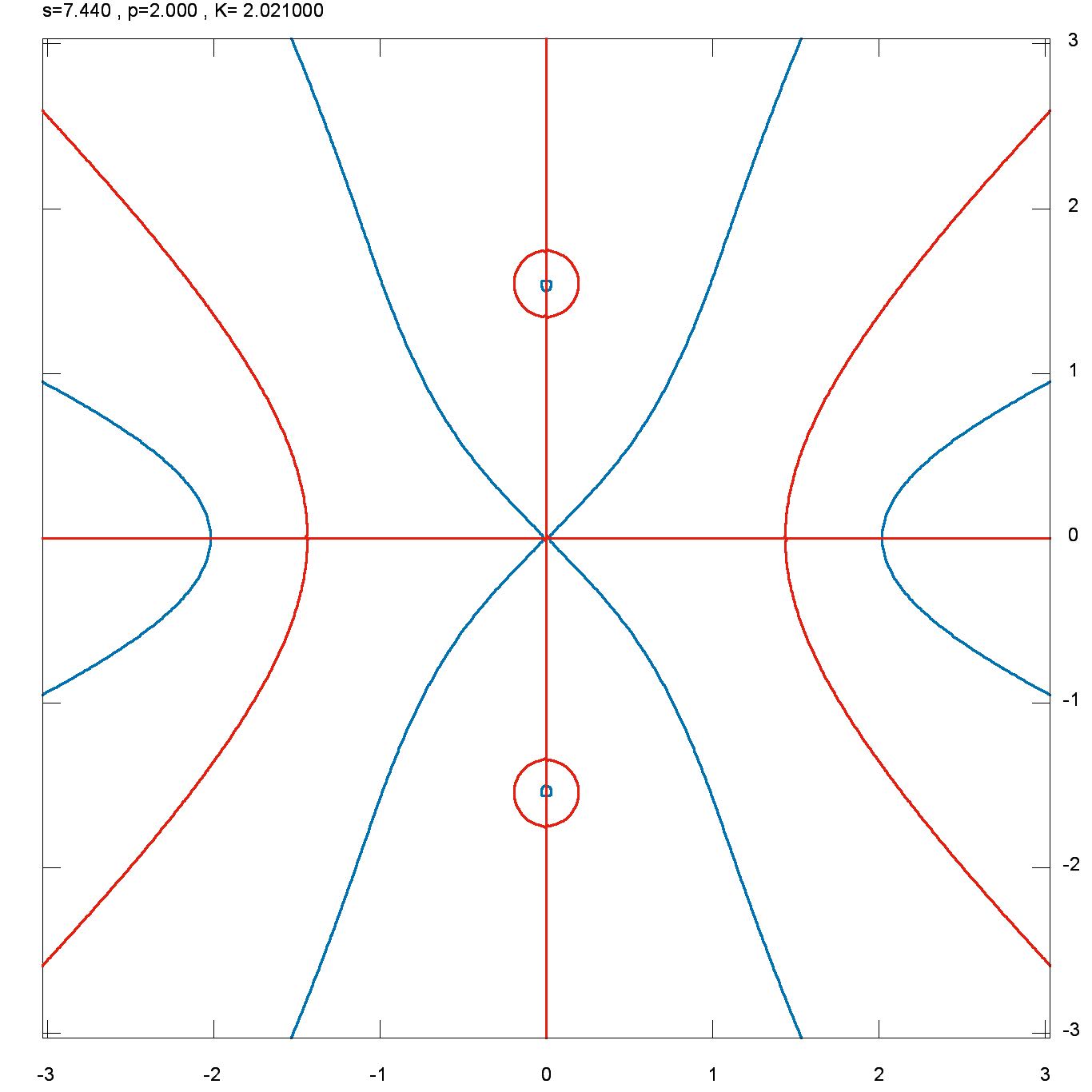}
}
\subfloat[$s$=7.44, $p$=2.0 and $K$=2.050\label{m:e}]{
  \includegraphics[width=45mm]{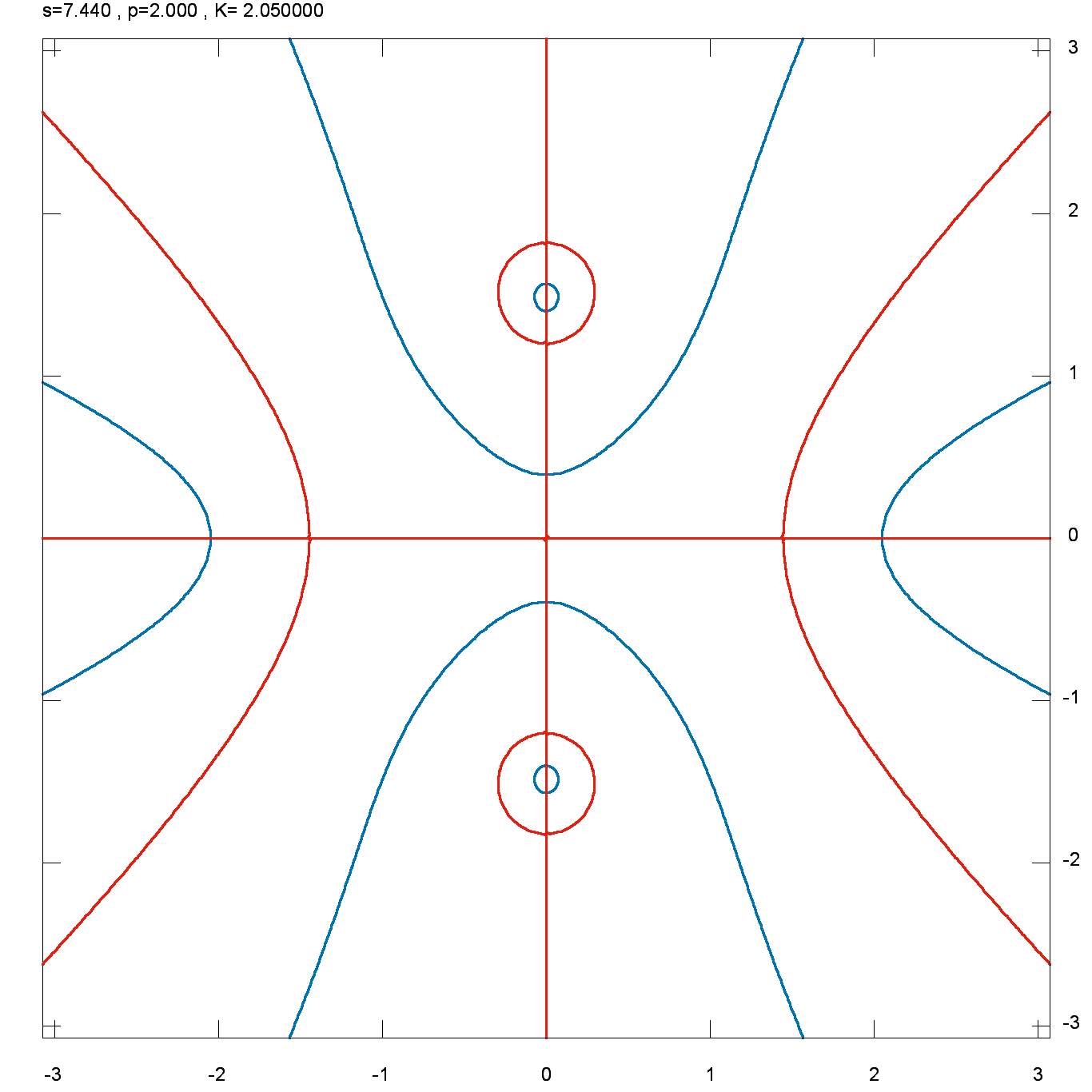}
}
\newline
\newline
\newline
\subfloat[$s$=7.44, $p$=2.0 and $K$=2.100\label{m:f}]{
  \includegraphics[width=45mm]{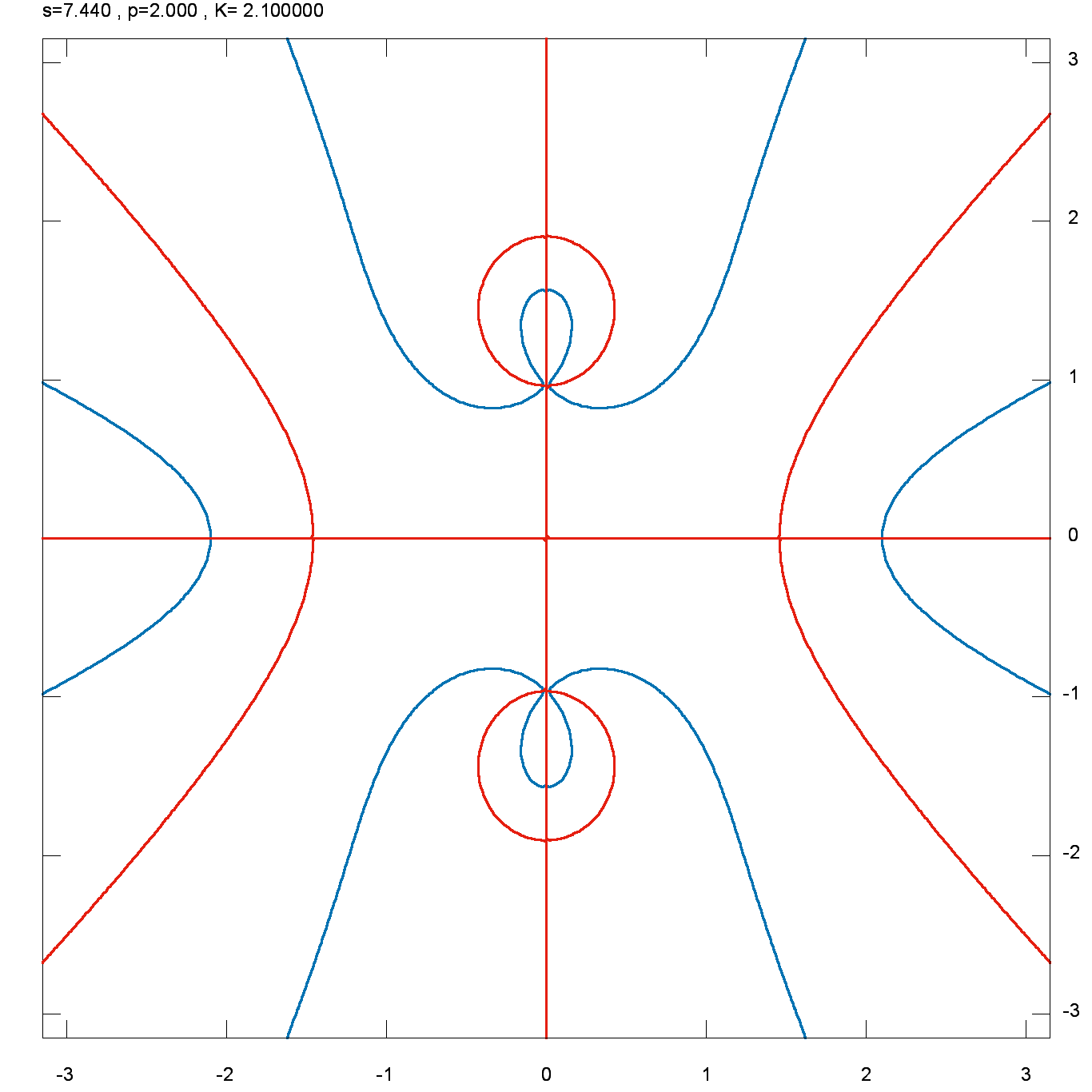}
}
\subfloat[$s$=7.44, $p$=2.0 and $K$=2.200\label{m:g}]{
  \includegraphics[width=45mm]{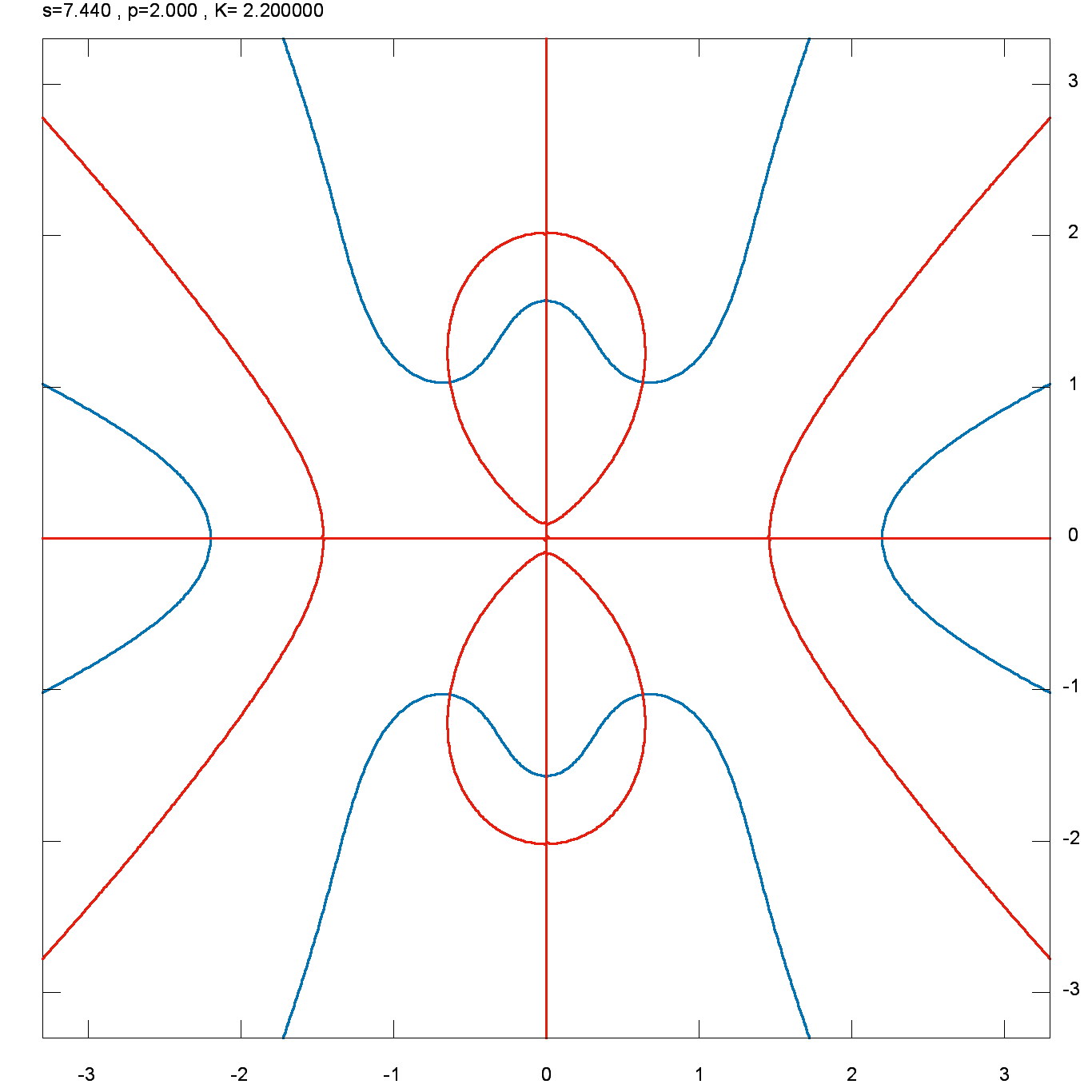}
}
\subfloat[$s$=7.44, $p$=2.0 and $K$=3\label{m:h}]{
  \includegraphics[width=45mm]{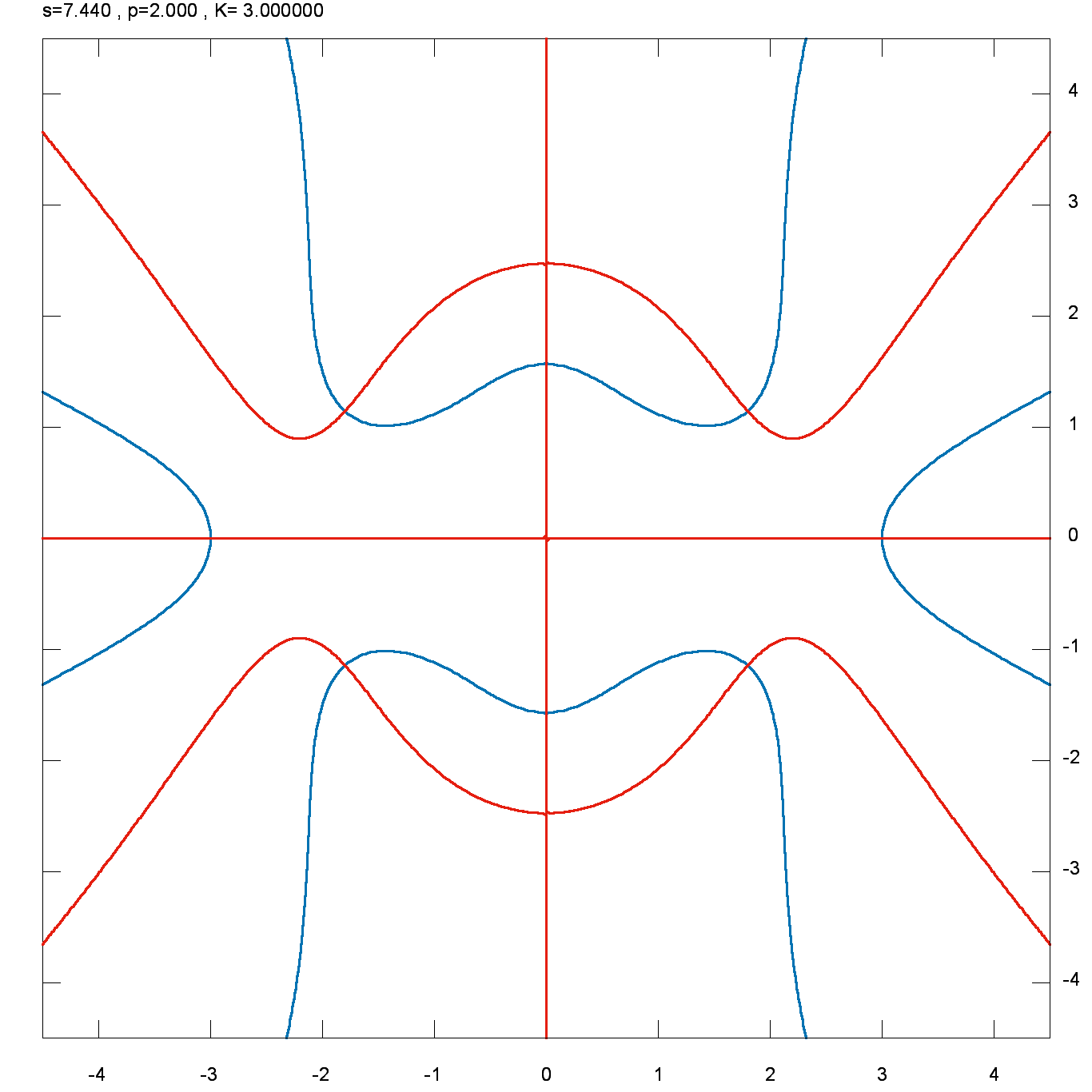}
}
\newline
\newline
\newline
\subfloat[$s$=7.44, $p$=2.0 and $K$=6\label{m:j}]{
  \includegraphics[width=45mm]{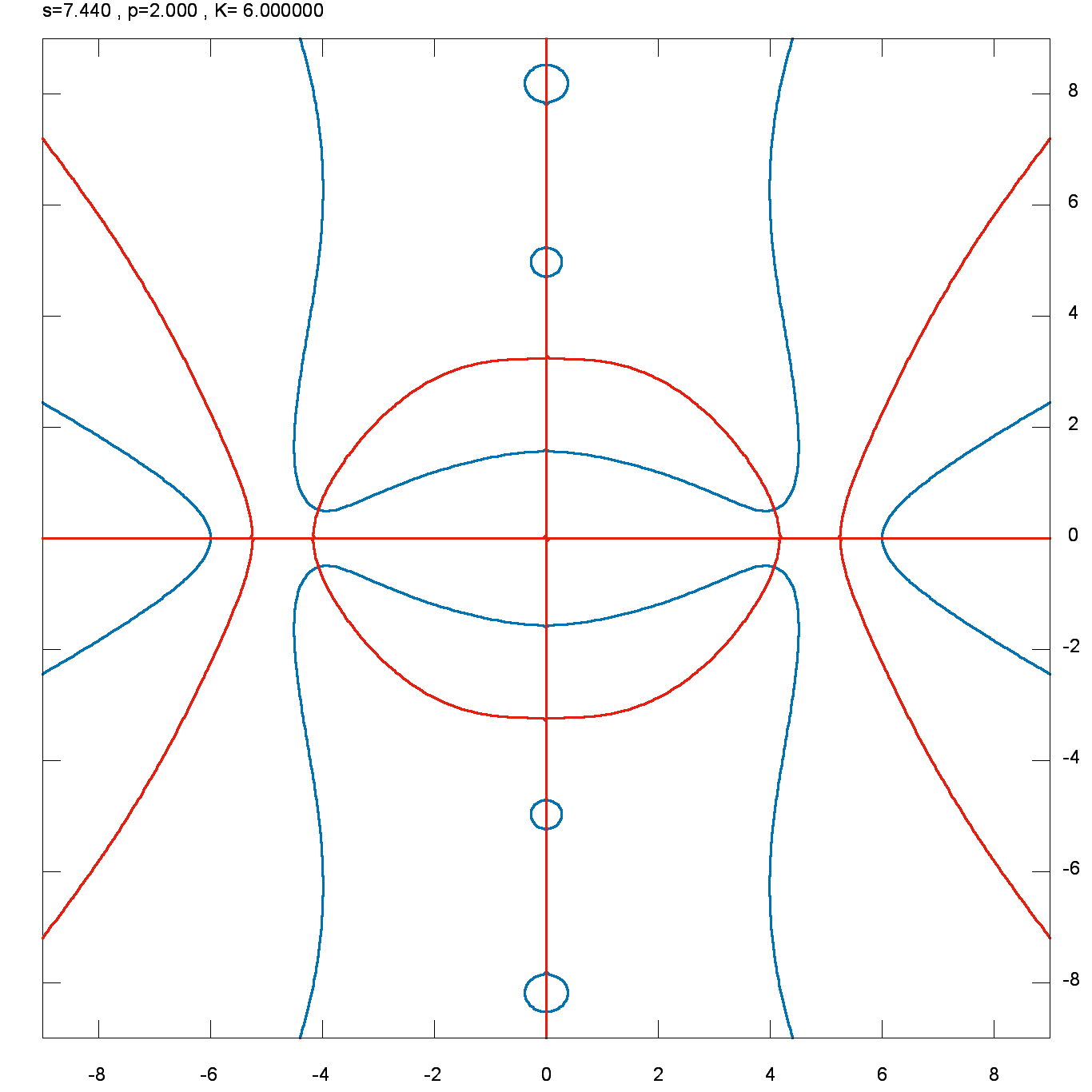}
}
\subfloat[$s$=7.44, $p$=2.0 and $K$=6.370\label{m:k}]{
  \includegraphics[width=45mm]{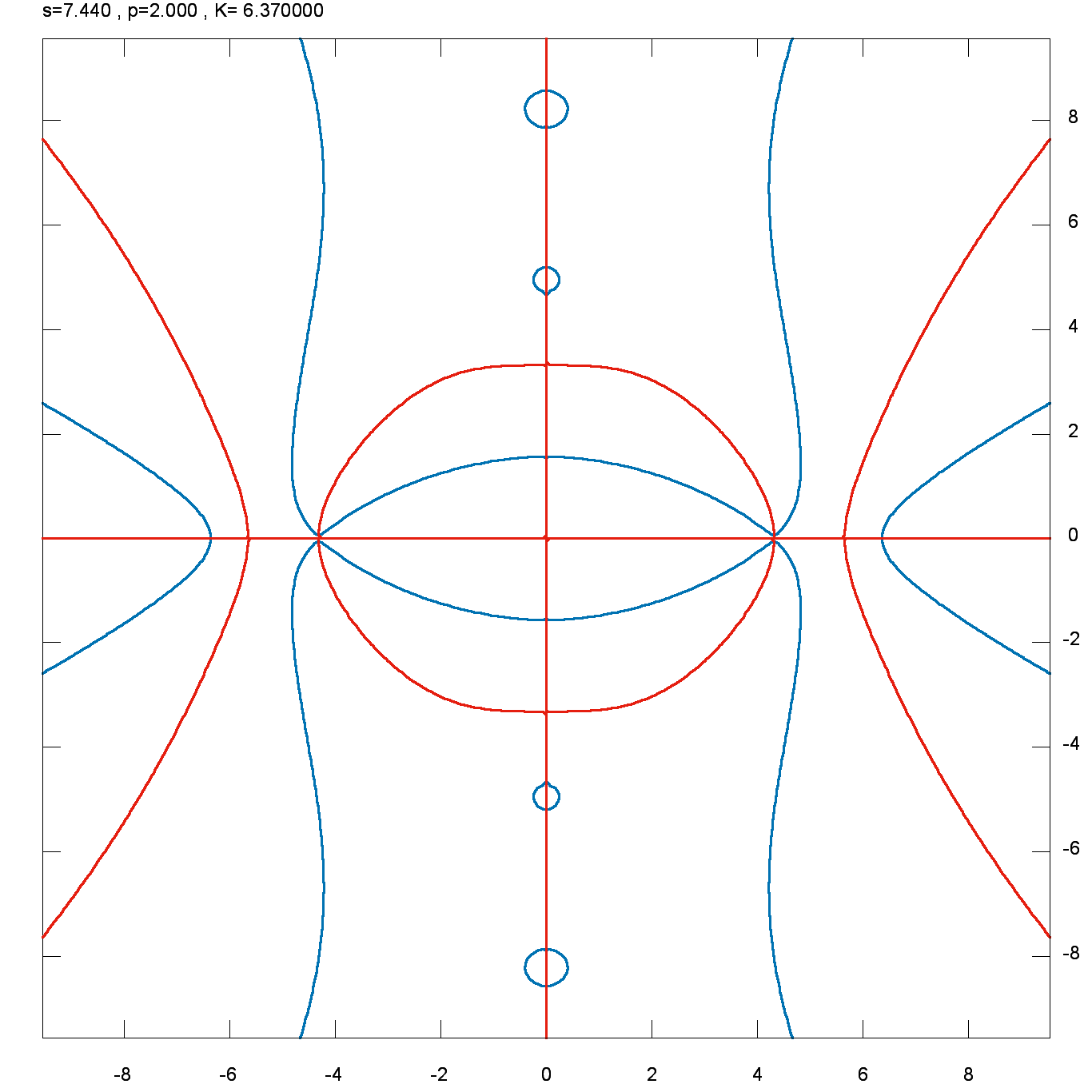}
}
\subfloat[$s$=7.44, $p$=2.0 and $K$=7\label{m:l}]{
  \includegraphics[width=45mm]{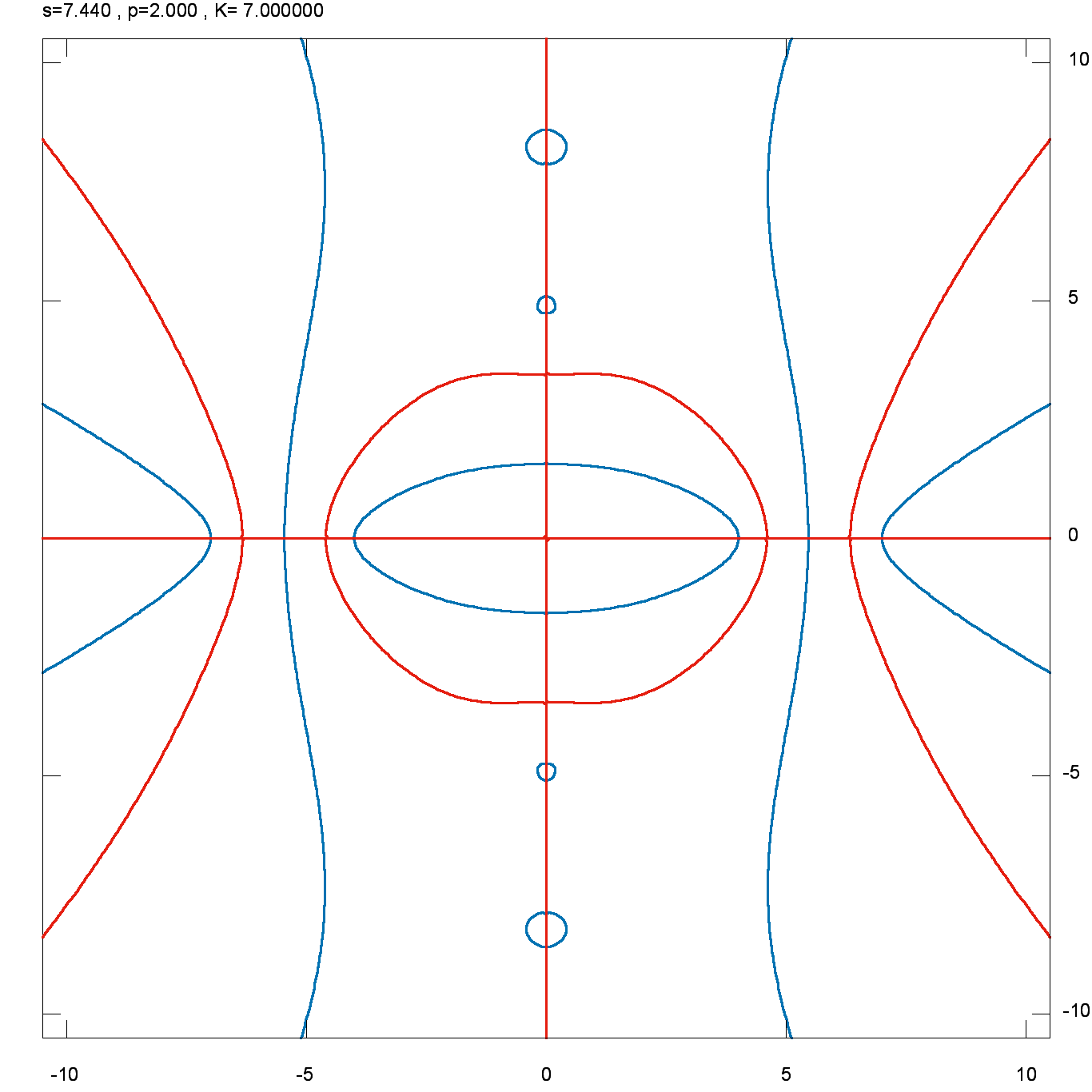}
}
\newline
\newline
\caption{Zero level lines of the real (blue) and imaginary (red) part of
  Eq.(\ref{e25}) plotted on the complex $Q$ plane at parameter values $p=2.0$
  and $s=7.44$. }
\vspace{-2pt}\label{qmod1}
\end{figure}

If, at a given $s$ and $p$ parameters one adjusts $K$, the type and number of
solutions changes. This is shown in Figs.\ref{m:c}-\ref{m:l} where
$s=7.44$ and $p=2.0$.\footnote{For technical reasons Eq.(\ref{e25}) was
  divided by $K^2 Q \cosh(K)\cosh(Q)$ and the result was plotted.} In Fig.\ref{m:c} one can see a single real (nontrivial)
solution. This transforms through a
bifurcation point (cf. Fig. \ref{m:d})  to an
imaginary solution, shown in Fig. \ref{m:e}.

\clearpage

\begin{figure}[!ht]
\centering
\subfloat[$s$=7.44, $p$=2.0 and $K$=1.0\label{n:a}]{
  \includegraphics[width=45mm]{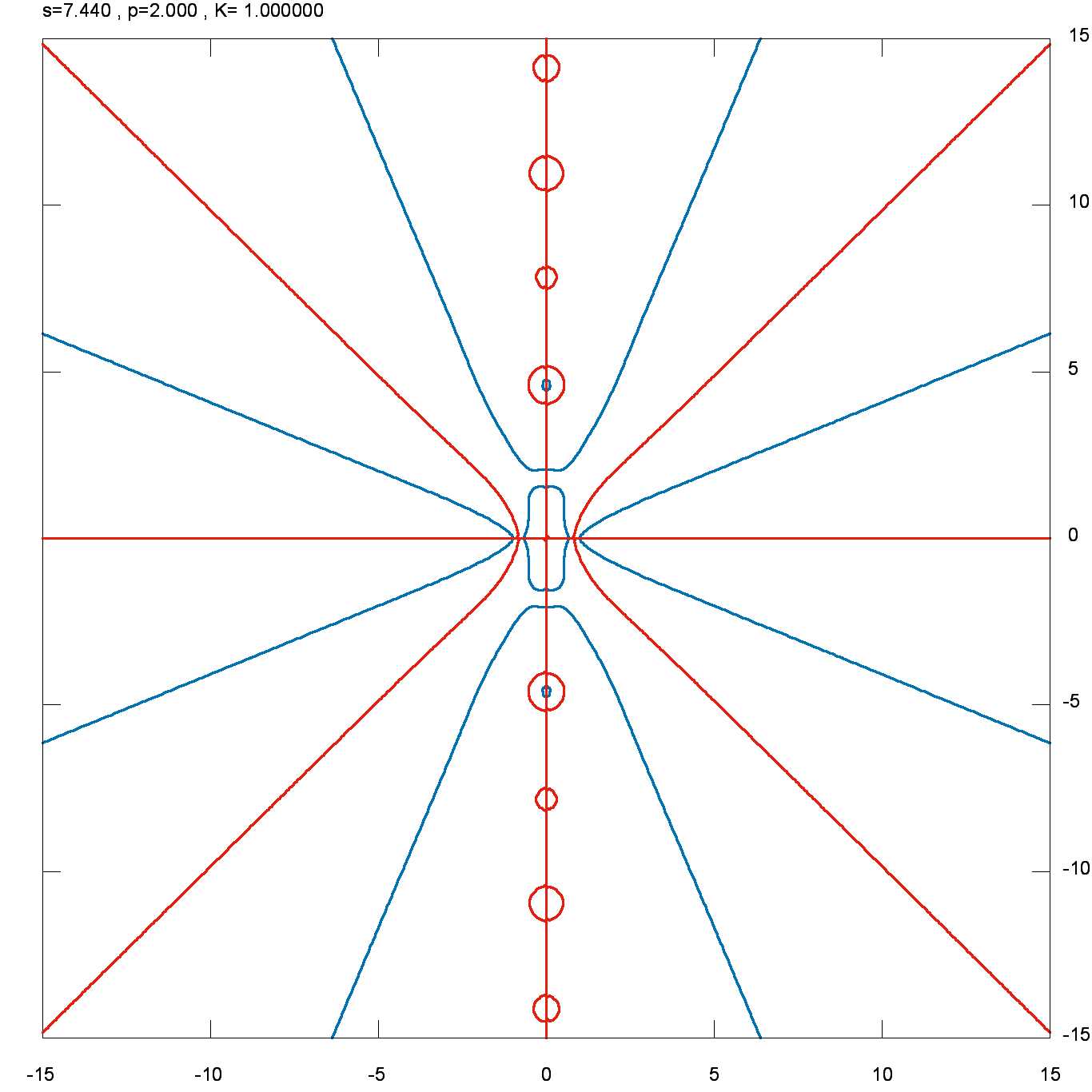}
}
\subfloat[$s$=7.44, $p$=2.0 and $K$=3.0\label{n:b}]{
  \includegraphics[width=45mm]{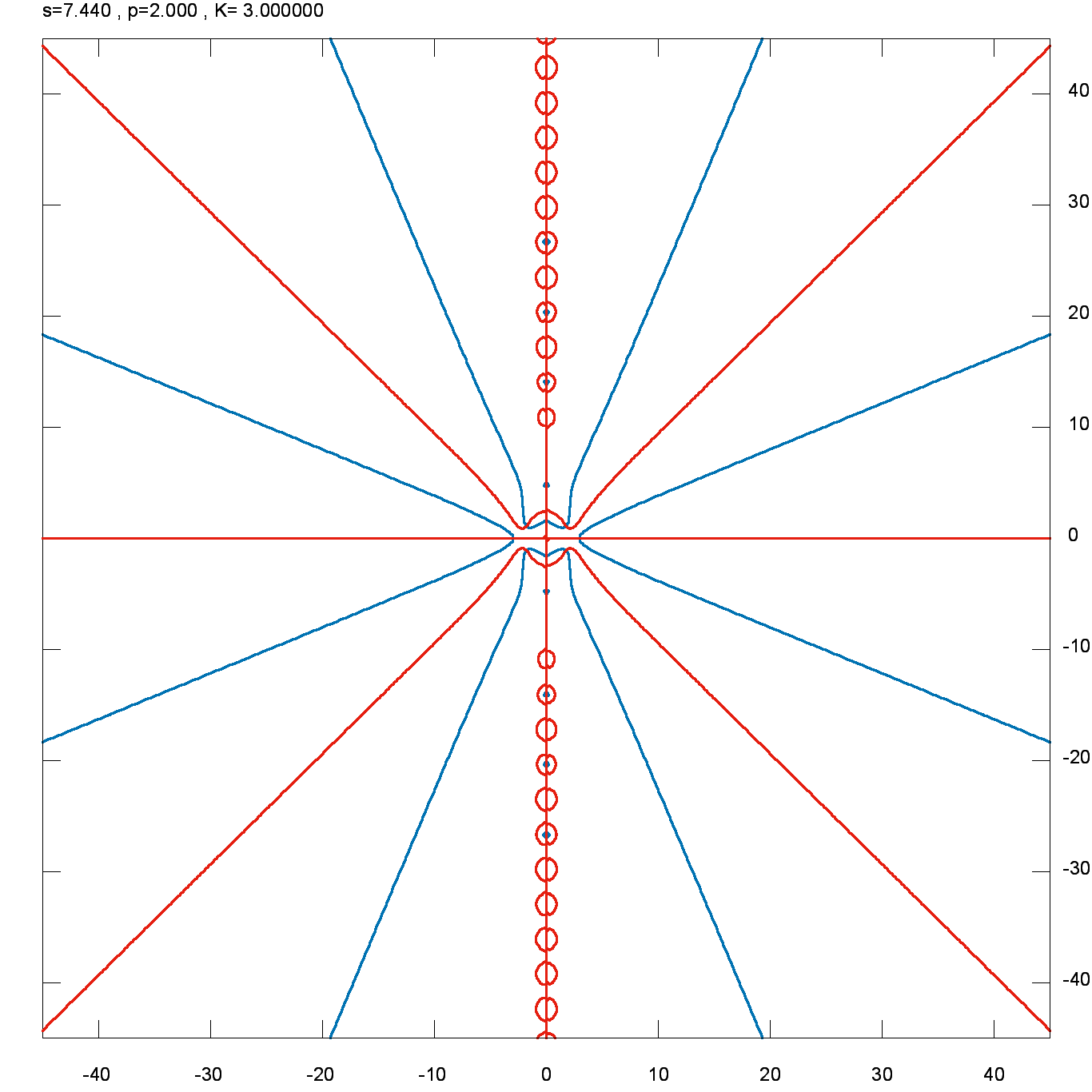}
}
\subfloat[$s$=7.44, $p$=2.0 and $K$=5.0\label{n:c}]{
  \includegraphics[width=45mm]{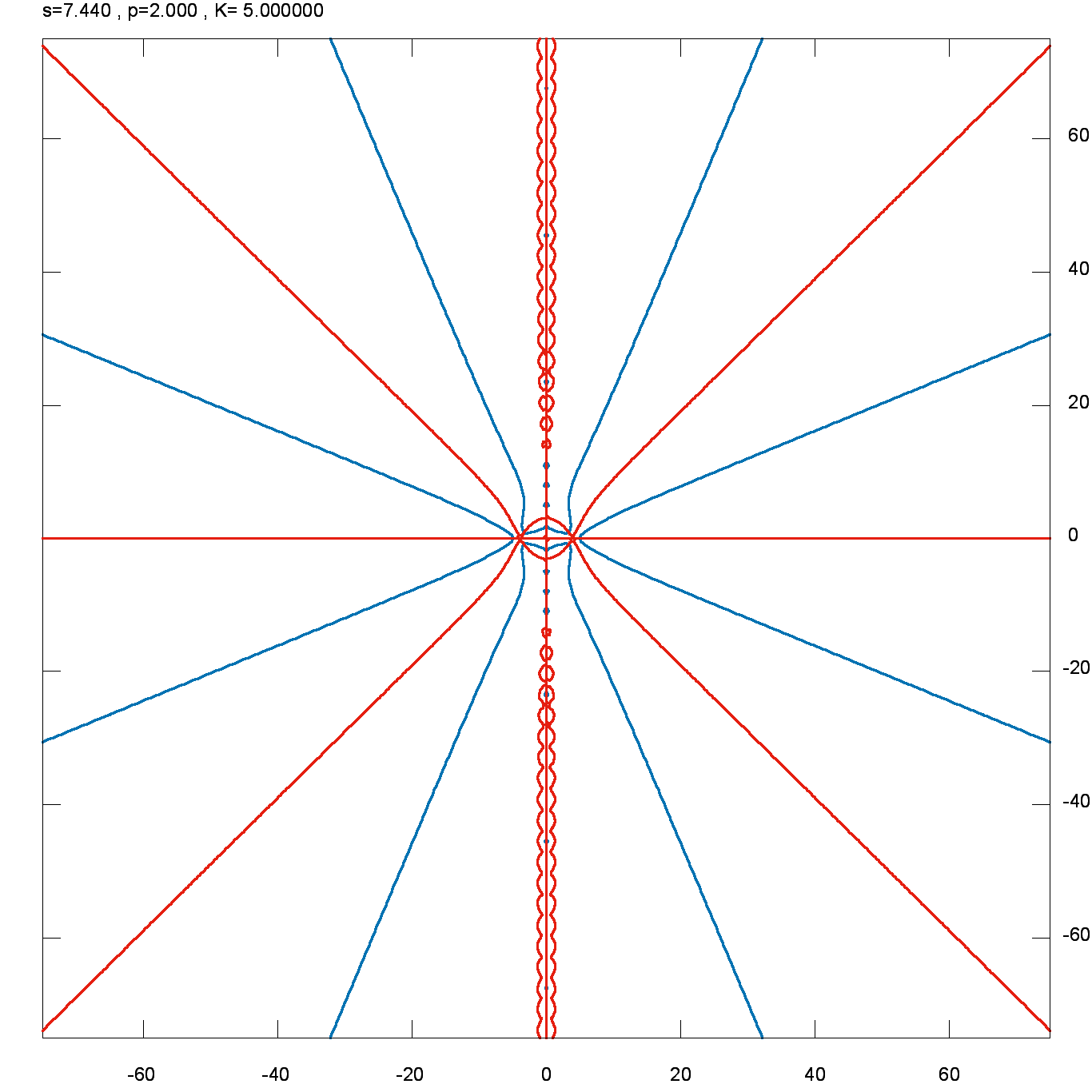}
}
\newline
\newline
\caption{Zero level lines of real and imaginary parts of Eq.(\ref{e25}) on the
  complex $Q$ plane. Several solutions along the imaginary axis are
  shown. Note that they are roughly equidistant and independent of the parameters, but the scales of the figures
  are different.
 } \label{qmod2}
\vspace{-2pt}
\end{figure}

Note that, as shown above, imaginary solutions always exist. This can be seen at larger scales
in Figs.\ref{n:a}-\ref{n:c}. 


Increasing the scaled wave number $K$
further, one can observe that two imaginary solutions collide
(see. Fig. \ref{m:f}) and give rise to a complex solution (see
Fig. \ref{m:g}). This complex solution gradually goes down to the real axis
(cf. Figs.\ref{m:g}-\ref{m:k}) and decays to two real solutions
(Fig.\ref{m:l}) which survive any further increase of $K$.

\begin{figure}[!ht]
\centering
\subfloat[$s$=7.44, $p$=3.0 and $K$=0.1\label{k:a}]{
  \includegraphics[width=45mm]{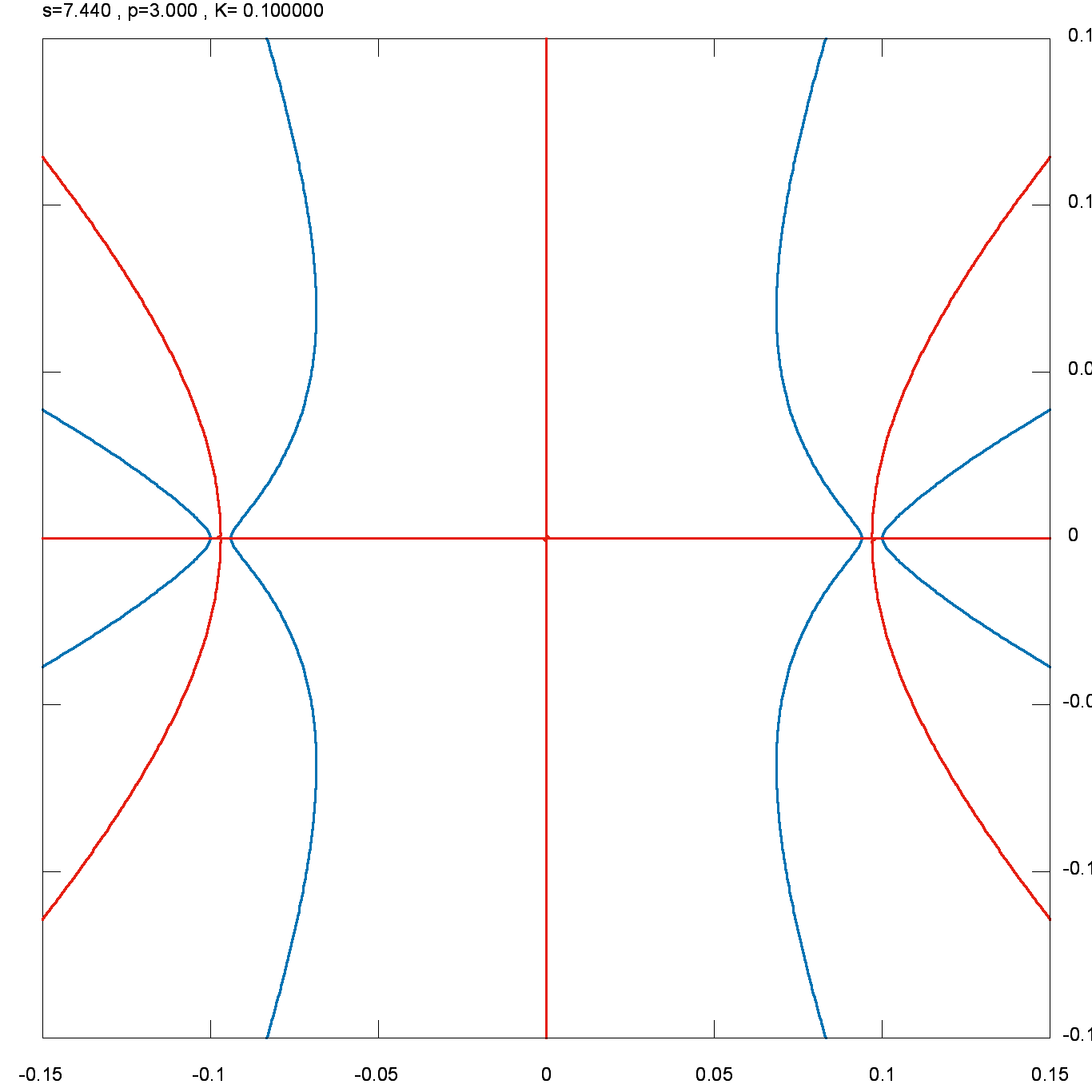}
}
\subfloat[$s$=7.44, $p$=3.0 and $K$=0.5\label{k:b}]{
  \includegraphics[width=45mm]{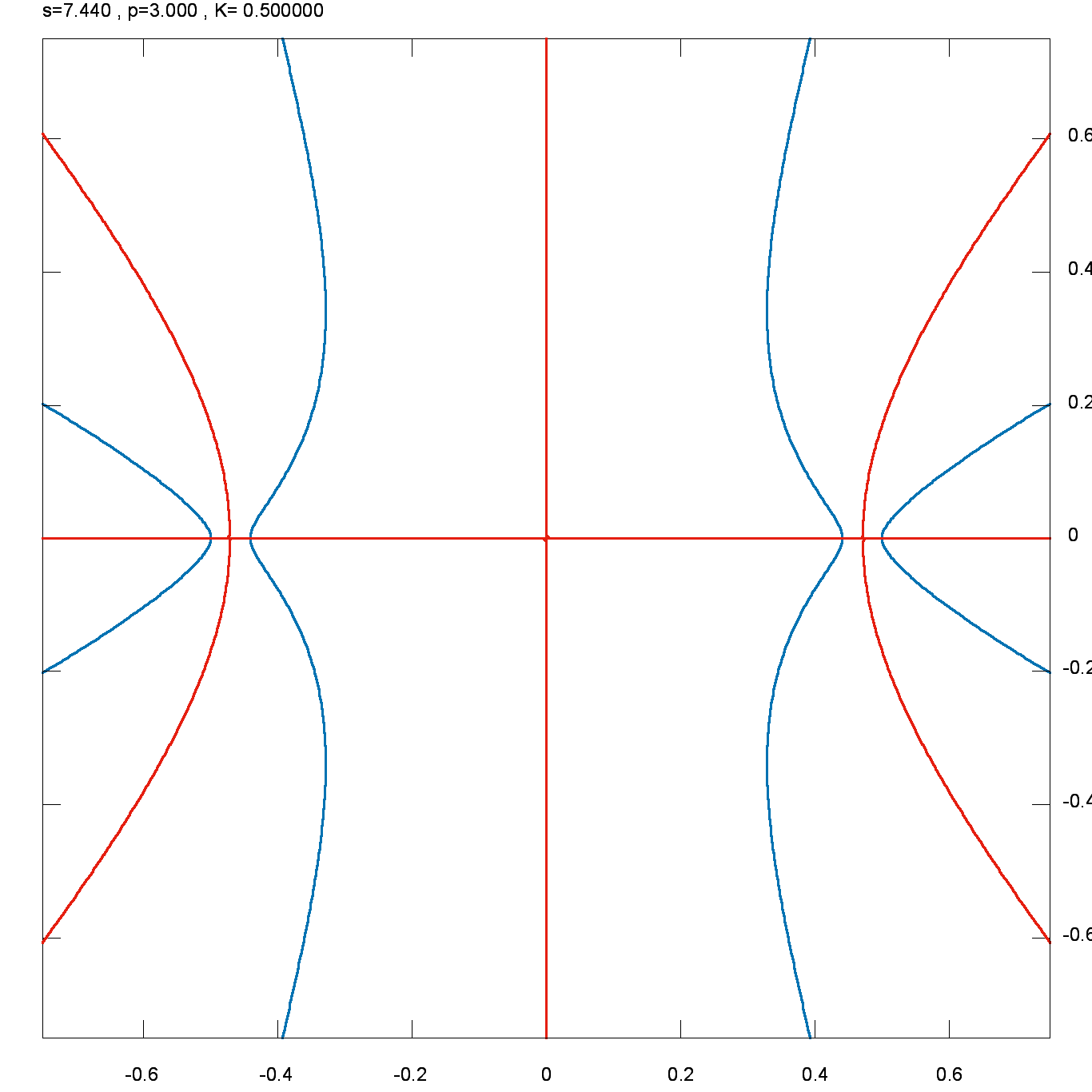}
}
\subfloat[$s$=7.44, $p$=3.0 and $K$=1.0\label{k:c}]{
  \includegraphics[width=45mm]{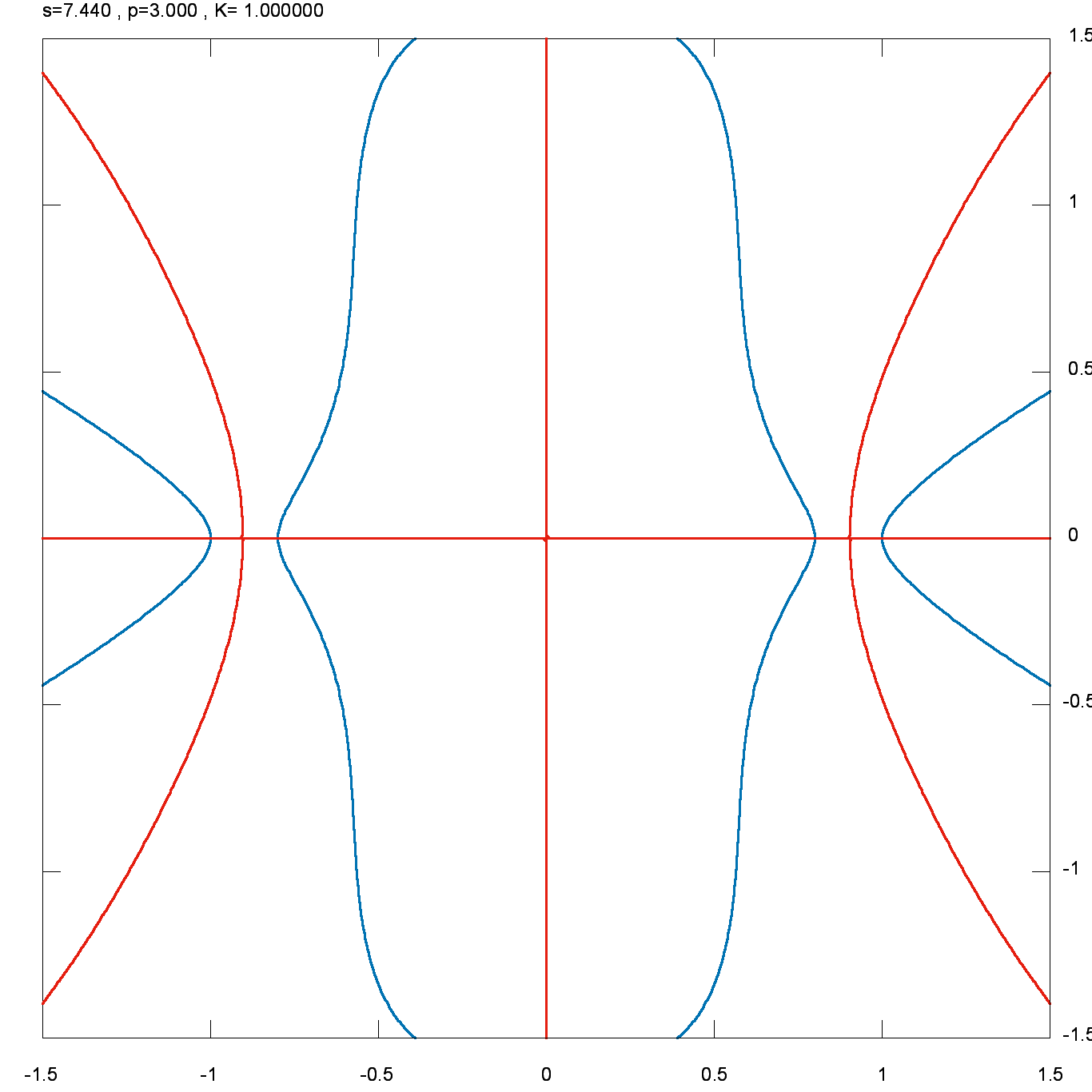}
}
\newline
\newline
\newline
\subfloat[$s$=7.44, $p$=3.0 and $K$=2.0\label{k:d}]{
  \includegraphics[width=45mm]{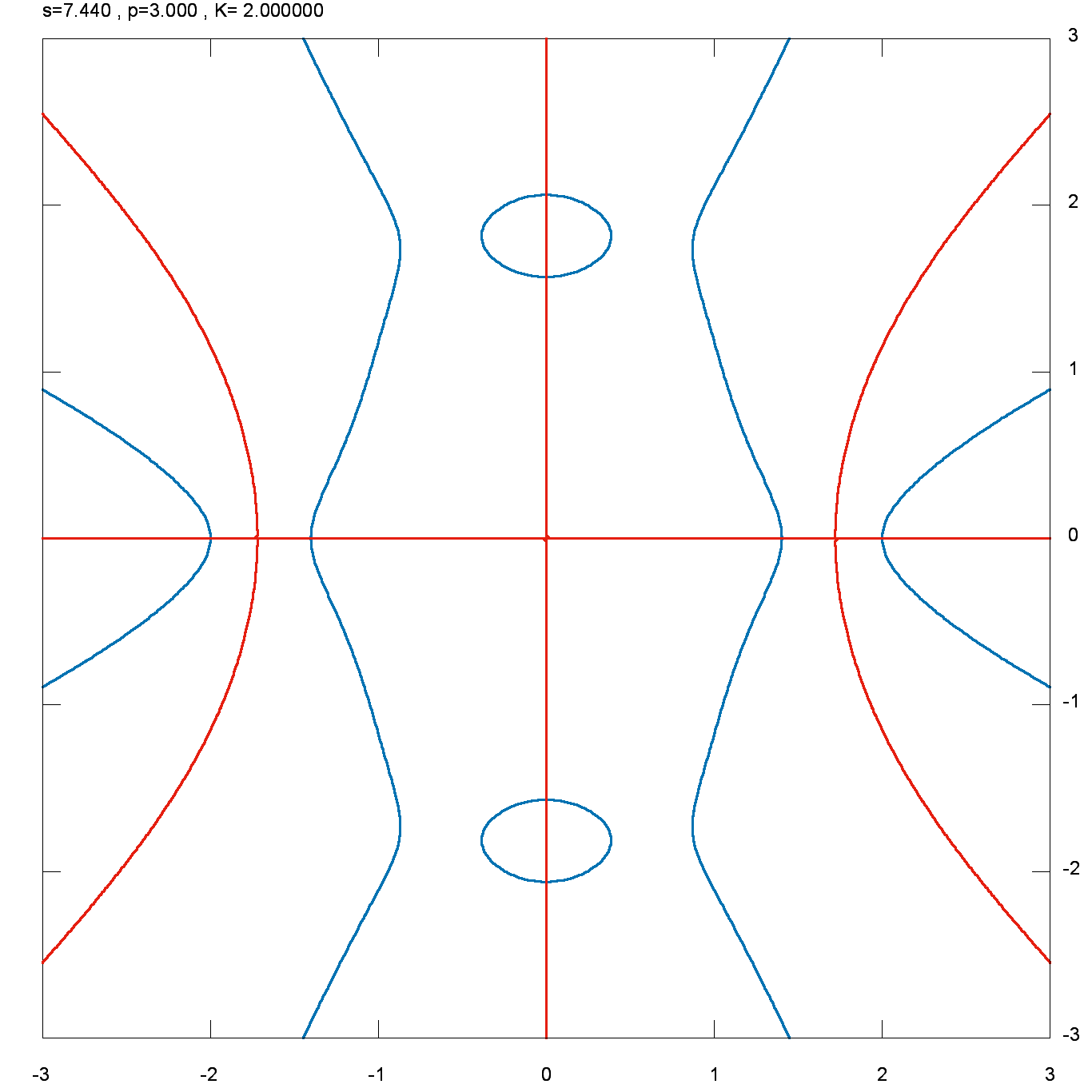}
}
\subfloat[$s$=7.44, $p$=3.0 and $K$=2.900\label{k:e}]{
  \includegraphics[width=45mm]{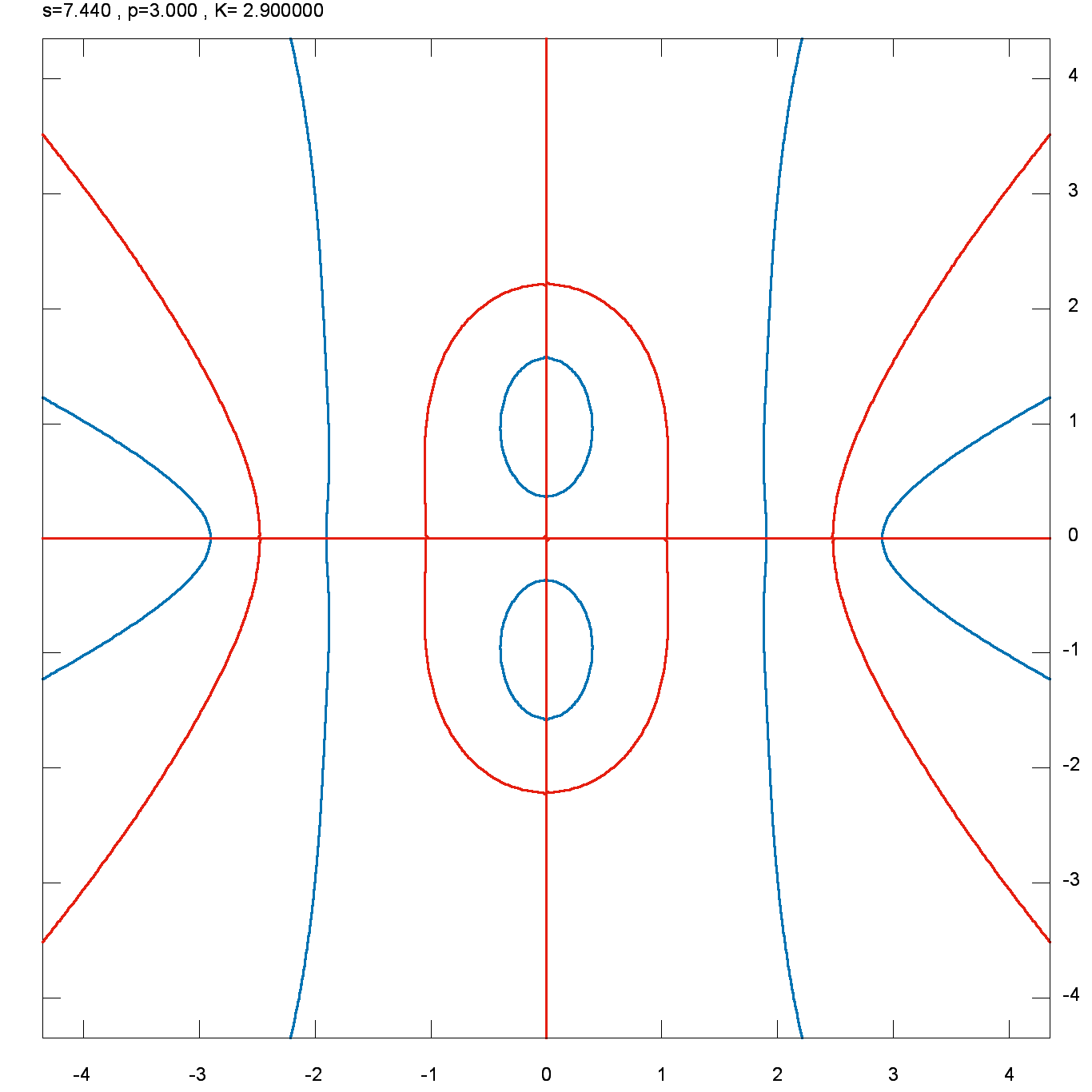}
}
\subfloat[$s$=7.44, $p$=3.0 and $K$=2.926\label{k:f}]{
 \includegraphics[width=45mm]{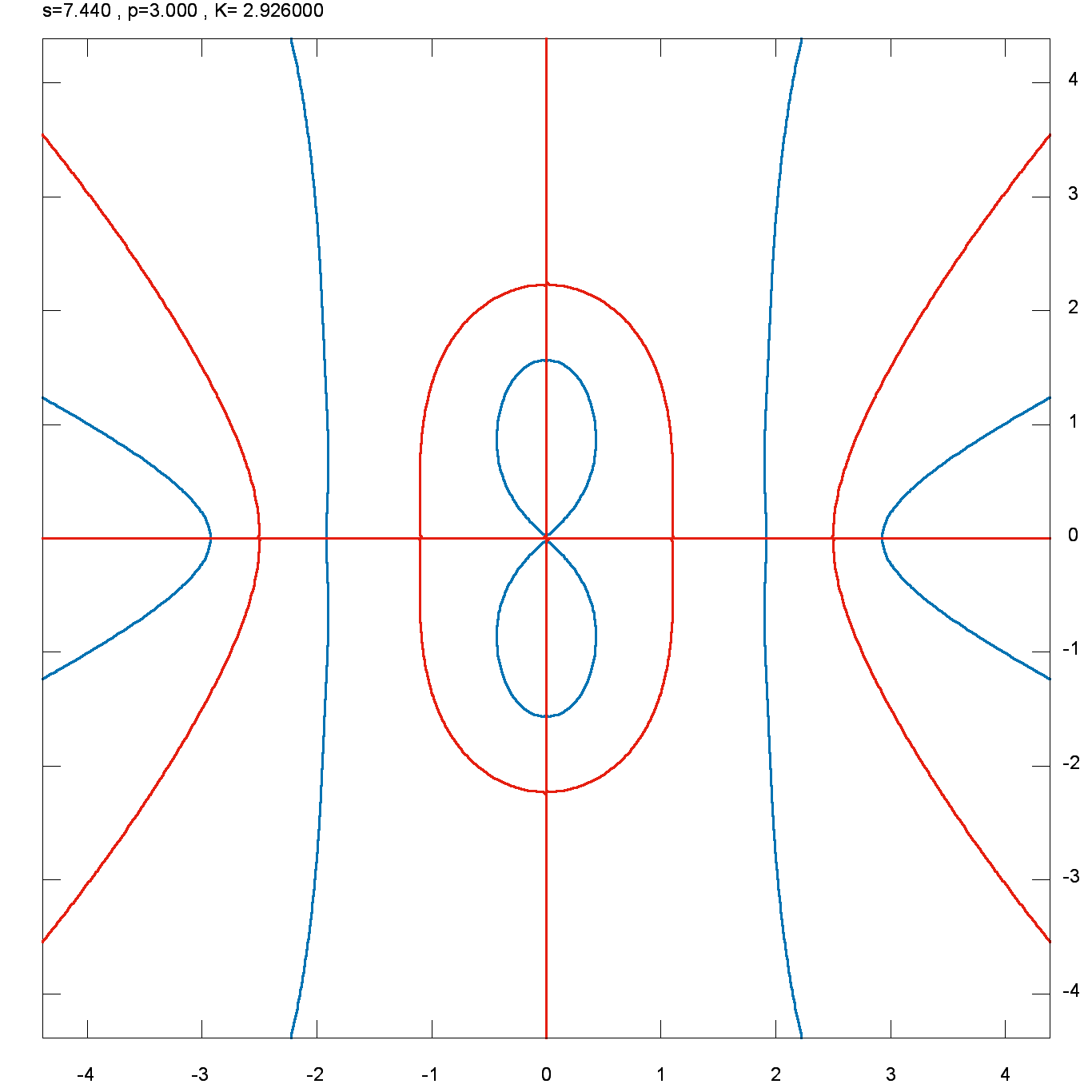}
}
\newline
\newline
\newline
\subfloat[$s$=7.44, $p$=3.0 and $K$=3\label{k:g}]{
  \includegraphics[width=45mm]{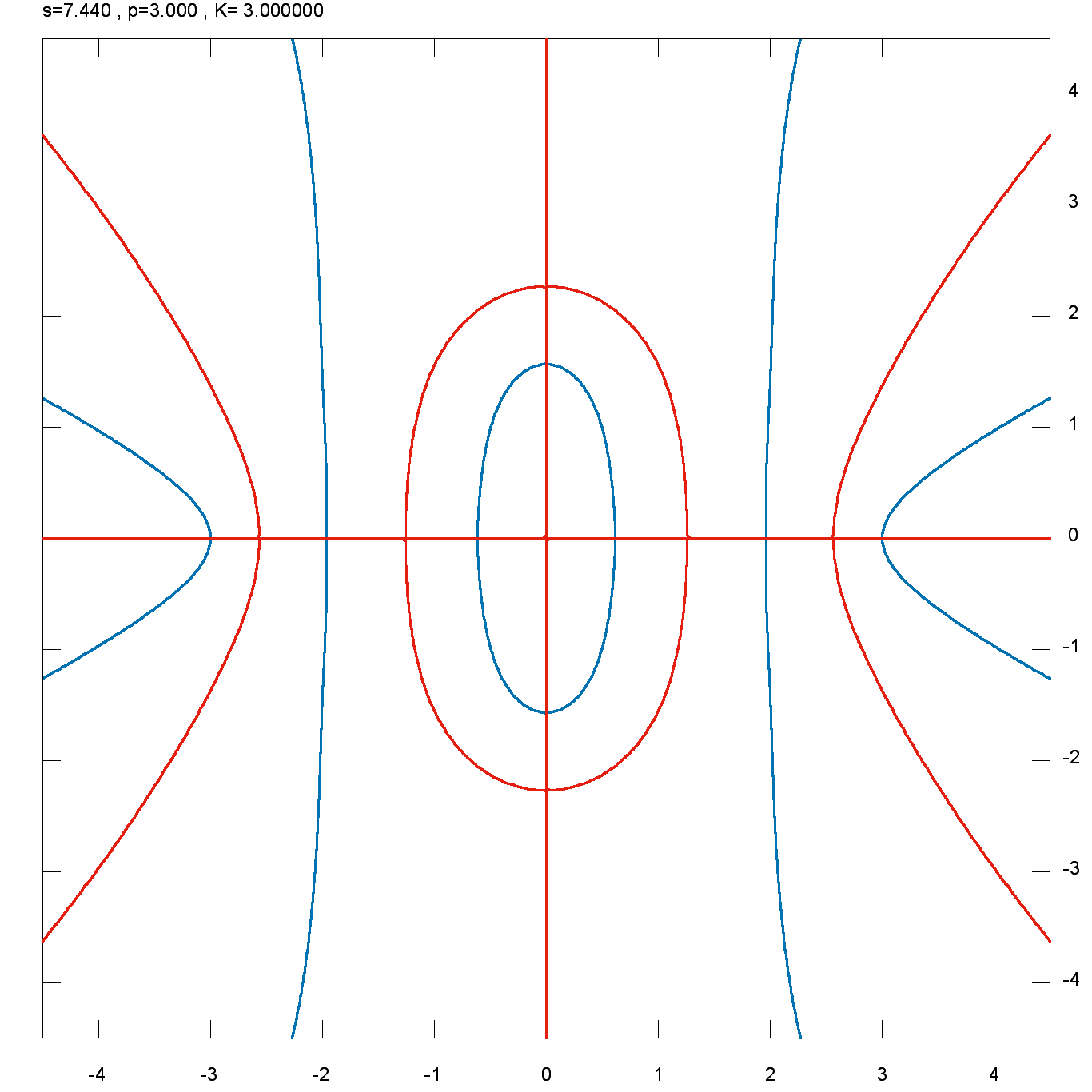}
}
\subfloat[$s$=7.44, $p$=3.0 and $K$=5\label{k:i}]{
  \includegraphics[width=45mm]{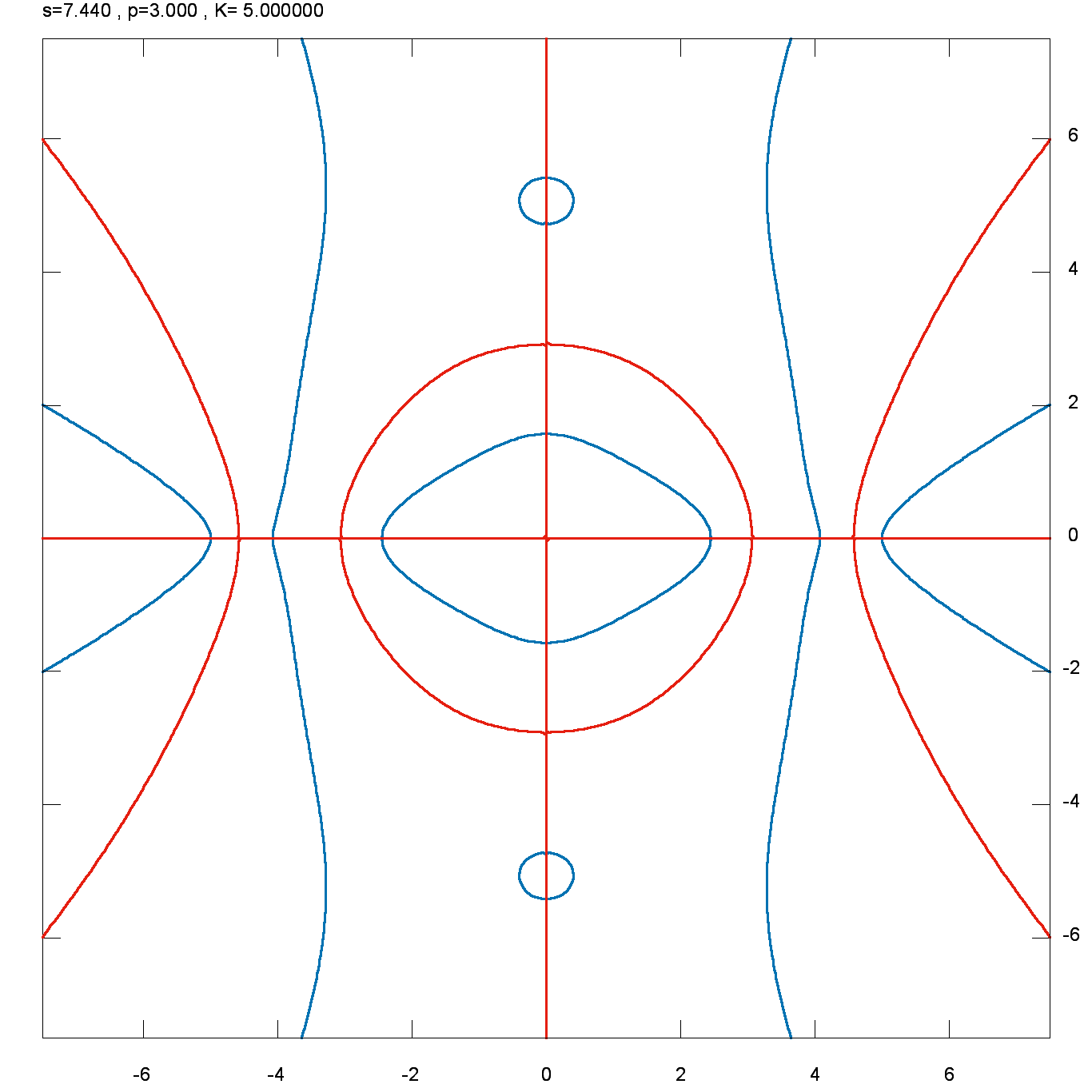}
}
\subfloat[$s$=7.44, $p$=3.0 and $K$=10\label{k:l}]{
  \includegraphics[width=45mm]{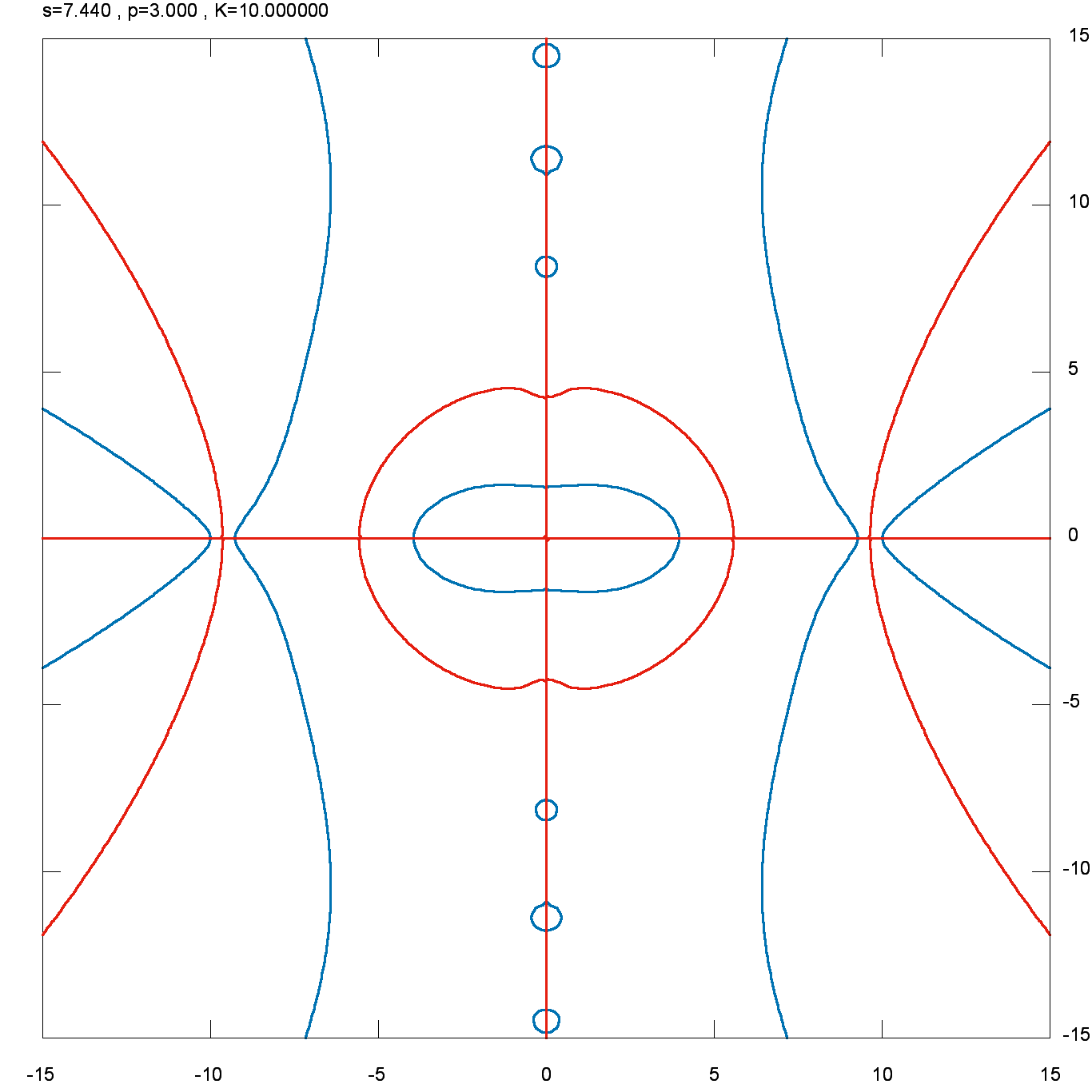}
}
\newline
\newline
\caption{Zero level lines of the real (blue) and imaginary (red) part of
  Eq.(\ref{e25}) plotted on the complex $Q$ plane at parameter values $p=3.0$
  and $s=7.44$. } \label{qmod3}
\end{figure}

Choosing the parameter values $s=7.44$ and $p=3.0$, one has a different
scenario, see Figs.\ref{k:a}-\ref{k:l}. This time only a single bifurcation
takes place, namely, an imaginary
solution and its mirror image collide at the origin (see \ref{k:f}) and
give rise to a second real solution (and its mirror image). 

\clearpage

In all these cases of birth of new solutions one can observe that approaching the
critical parameter the zero level lines of the real part of
Eq.(\ref{e25}) develop an edge, and at bifurcation they become (locally) two 
straight lines crossing each other on the real and/or the imaginary axis. This
means that at the bifurcation point the derivative of
the line is undetermined, i.e., has a form $\frac{0}{0}$. Note in passing that it is
equivalent with the condition that the second derivative becomes infinite.
Let us denote for brevity $\Re(Q)$ with $\alpha$ and $\Im(Q)$ with $\beta$, further, the
left hand side of Eq.(\ref{e25}), divided by $K^2 Q \cosh(K)\cosh(Q)$ be $D(\alpha+i\beta)$. Then the zero level line  of 
the real part of $D$ is given by
 \begin{eqnarray}
\Re\left(D(\alpha+i\beta)\right)=0\;.
\label{mm1}
\end{eqnarray}
Taking its derivative with respect to $u$, we get
\begin{eqnarray}
\Re\left(D'(\alpha+i\beta)+iD'(\alpha+i\beta)\frac{d\beta}{d\alpha}\right)=0\;,
\label{mm2}
\end{eqnarray}
hence the derivative of the curve is written as
\begin{eqnarray}
\frac{d\beta}{d\alpha}=\frac{\Re\left(D'(\alpha+i\beta)\right)}{\Im\left(D'(\alpha+i\beta)\right)}\;.
\label{mm3}
\end{eqnarray}
It follows that at bifurcation
\begin{eqnarray}
D'(\alpha+i\beta)=0
\label{mm5}
\end{eqnarray}
must be satisfied, together with $D(\alpha+i\beta)=0$. Now it is easily seen that
$D(\alpha+i\beta)$ is real along both the real and the imaginary axis, while
$D'(\alpha+i\beta)$ is real along the real axis and imaginary  along the imaginary
axis. Therefore, in case of a bifurcation on the real axis we have
\begin{eqnarray}
&&D(\alpha)=0\label{mm5a1}\\
&&D'(\alpha)=0\;,
\label{mm5a2}
\end{eqnarray}
i.e., two real equations for the two real parameters $u$ and $p$. 
Similarly, in case of a bifurcation on the imaginary axis we have
\begin{eqnarray}
&&D(i\beta)=0\label{mm5c1}\\
&&iD'(i\beta)=0\;,
\label{mm5c2}
\end{eqnarray}
again two real equations for two real parameters.

Eq.(\ref{e25}) shows that $D(Q)$ may be expressed as
\begin{eqnarray}
D(Q)=\left(1+sK^2\right)F(Q)+p\;G(Q)
\label{mm5aa}
\end{eqnarray}
where
\begin{align} \label{mm5ab}
F(Q) & = \frac{\tanh K}{K}-\frac{\tanh Q}{Q}\;,\\
G(Q) = & -4\frac{K^2+Q^2}{\cosh K \cosh
    Q}+\left(\frac{Q^4}{K^2}+2Q^2+5K^2\right) \\ 
& -\left( Q^4+6K^2Q^2+K^4\right) \frac{\tanh K}{K}\frac{\tanh Q}{Q} \nonumber.
\end{align}
In terms of these functions we have (cf. Eqs.(\ref{mm5a1}), (\ref{mm5a2}))
\begin{align}
&F(\alpha)'(\alpha)-G(\alpha)F'(\alpha)=0\label{mm5b1}\\
&\frac{p}{1+sK^2}=-\frac{F(\alpha)}{G(\alpha)}
\label{mm5b2}
\end{align}
and (cf. Eqs.(\ref{mm5c1}), (\ref{mm5c2}))
\begin{align}
&a(i\beta)G'(i\beta)-G(i\beta)F'(i\beta)=0\label{mm5d1}\\
&\frac{p}{1+sK^2}=-\frac{F(i\beta)}{G(i\beta)}
\label{mm5d2}
\end{align}
for bifurcations on the real and imaginary axis, respectively. In these cases
Eq.(\ref{mm5b1}) or Eq.(\ref{mm5d1}) is solved numerically to get $\alpha$ or $\beta$,
respectively, then the solution is inserted into Eqs.(\ref{mm5b2}), (\ref{mm5d2}), respectively, which in turn are solved for \(p\), taking into account Eq.(\ref{e25f}).

\begin{figure}[H]
\centering
\includegraphics[width=12.5 cm]{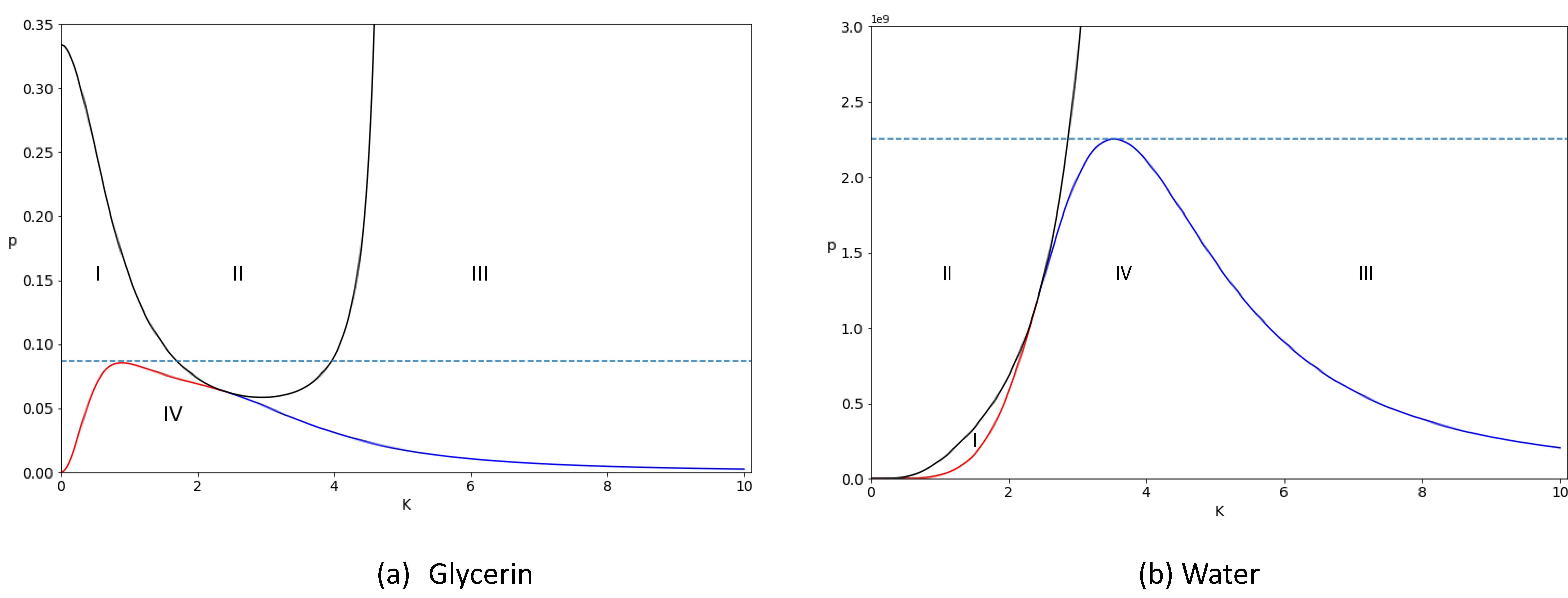}
\caption{The maximal parameters $p$ versus the scaled wave-number $K$ for glycerin (a) and water (b). The red line signifies the emergence of complex solutions at the imaginary line, while the blue line represents the real solutions obtained from Eqs. (\ref{mm5b1}) and (\ref{mm5b2}). The black line denotes the onset of the crossover from imaginary to real solutions at the origin.} \label{fig1}
\end{figure}

As for the transformation of an imaginary solution to a real one at the
origin, mentioned above, for the critical $p(K)$ line  one obtains
\begin{eqnarray}\frac{p}{1+sK^2}=-\frac{F(0)}{G(0)}=\frac{1-\frac{\tanh
      K}{K}}{5K^2-4\frac{K^2}{\cosh(K)}-K^3\tanh K}\;.
\label{mm5e}
\end{eqnarray}
Since both $F(Q)$ and $G(Q)$ are even functions of $Q$, it follows that
$F'(0)=G'(0)=0$, hence Eqs.(\ref{mm5b1}), (\ref{mm5d1}) are automatically
satisfied for $\alpha=0$ and $\beta=0$, respectively. 
The results are plotted in
Figs.\ref{fig1}, \ref{fig2}. 

In the figures red line means the onset of creation of complex solution at the
imaginary line, i.e., the solution of Eqs.(\ref{mm5d1}), (\ref{mm5d2}). Blue line is the
same at the real line (cf. Eqs.(\ref{mm5b1}), (\ref{mm5b2})), while black line is
the onset of crossover from imaginary to real solution at the origin (Eq.(\ref{mm5e})). It is obvious that these lines must have a common point. The lines partition the $K-p$ parameter space to four regions, denoted in the figures by Roman numbers: 

\begin{enumerate}[I]
\item Only imaginary solutions (infinitely many of them) are present.
\item There is a single real solution and there are infinitely
  many imaginary solutions.
\item There are two real solutions and there are infinitely
  many imaginary solutions.
\item There is a single complex solution and there are infinitely many imaginary solutions.
\end{enumerate}

Asymptotics and some special cases of the bifurcation curves  are the following.
\begin{enumerate}
\item Imaginary to complex $Q$ (border between regions I. and IV., red curve in Fig.\ref{fig1} ):

For small $K$ ($K\ll 1$) we have
\begin{eqnarray}
&&Q=(1.1127+0.2509\;K^2)i\\
&&p=0.53667\;K^2\left(1+(s-3.8674)K^2\right)
\label{asy2a}
\end{eqnarray}
\item Complex to real $Q$ (border between regions IV. and III., blue curve in Fig.\ref{fig1} ):

For large $K$ we have
\begin{eqnarray}
&&Q=0.6823\;K\\
&&p=\frac{1.7200(1+sK^2)}{K^3}\;.
\label{asy2b}
\end{eqnarray}
\item Imaginary to real $Q$ (borders between regions I., II. and III., black curve in Fig.\ref{fig1} ):

For small $K$  we have
\begin{eqnarray}
p=\frac{1}{3}\left[1+\left(s-\frac{7}{5}\right)K^2\right]\;.
\label{asy3a}
\end{eqnarray}
At $K\approx 4.9435$ parameter $p$ diverges as
\begin{eqnarray}
p=\frac{0.8452+20.654\; s}{4.9435-K}\;,
\label{asy3b}
\end{eqnarray}
For a given material this implies (cf. Eq.\ref{e25f})
\begin{eqnarray}
p=\frac{8.8\times 10^3\; \mu^{6}}{\left(4.9435-K\right)^3}\;.
\label{asy3ba}
\end{eqnarray}
\item The common point of the bifurcation parameter curves (red, blue and
  black lines in  Fig. (\ref{fig1}) satisfies
\begin{eqnarray}
\frac{F(0)}{G(0)}=-\frac{p}{1+sK^2}=\lim_{Q\rightarrow 0}\frac{F'(Q)}{G'(Q)}=\frac{F''(0)}{G''(0)}\;,
\label{asy4a}
\end{eqnarray}
since $F'(0)=G'(0)=0$.
This implies 
\begin{eqnarray}
F(0)G''(0)-G(0)F''(0)=0\;.
\label{asy4b}
\end{eqnarray}
The solution of this equation yields for the coordinates of the common point
\begin{eqnarray}
&&K=2.4152\\
&&p=0.05307(1+5.8332\;s)\;.
\label{asy4c}
\end{eqnarray}
\end{enumerate}


\section{Minimal layer thickness necessary for wave propagation}

Viscosity not only damps waves, but it can even prevent their
propagation. Indeed, propagation, mathematically, a real part of the complex
angular frequency, appears only in region IV. (cf. Fig \ref{fig1}). This implies that
no gravity-capillary waves can propagate if \(p\) is large enough (or, equivalently, if the layer width is small enough.
Further, even if the layer thickness is larger than the critical value, neither very long, nor very short waves can propagate.

The critical layer thickness is found from the maximum point of the curves bordering region IV. in Fig.\ref{fig1} (cf. Eqs.(\ref{mm5b1})-(\ref{mm5d2})). This depends on material material parameter \(\mu\), so we present the results in Table.\ref{material_tbl}. Clearly, for water and mercury the critical layer width is so extremely small that at such scales even the applicability of standard hydrodynamics is more than questionable.

\begin{table}[]
\caption{material parameters $l_{\nu}$, $l_{\sigma}$ and $\mu$ for some fluids.\label{tab1}}
\centering
\begin{tabular}{cccccc}
 \hline Material & $l_{\nu}$ & $l_{\sigma}$ & $\mu$ & $p_{max}$ & $h$ \\ 
 \hline water		& $4.67\times 10^{-5}$	& $2.73\times 10^{-3}$  & 58.40 & $2.255\times 10^{9}$ & $3.561\times 10^{-8}$\\ 
 \hline glycerin    & $5.04 \times 10^{-3}$     & $2.27 \times 10^{-3}$   & 0.45 & 0.085 & $1.146 \times 10^{-2}$\\ 
 \hline mercury     & $1.09\times 10^{-5}$    & $1.90\times 10^{-3}$    & 174.86 & $1.625 \times 10^{12}$ & $9.271\times 10^{-10}$ \\ 
 \hline
\end{tabular}
\end{table}

\section{Particle Motion at Surface}
\begin{figure}[H]
\centering
\includegraphics[width=9.5 cm]{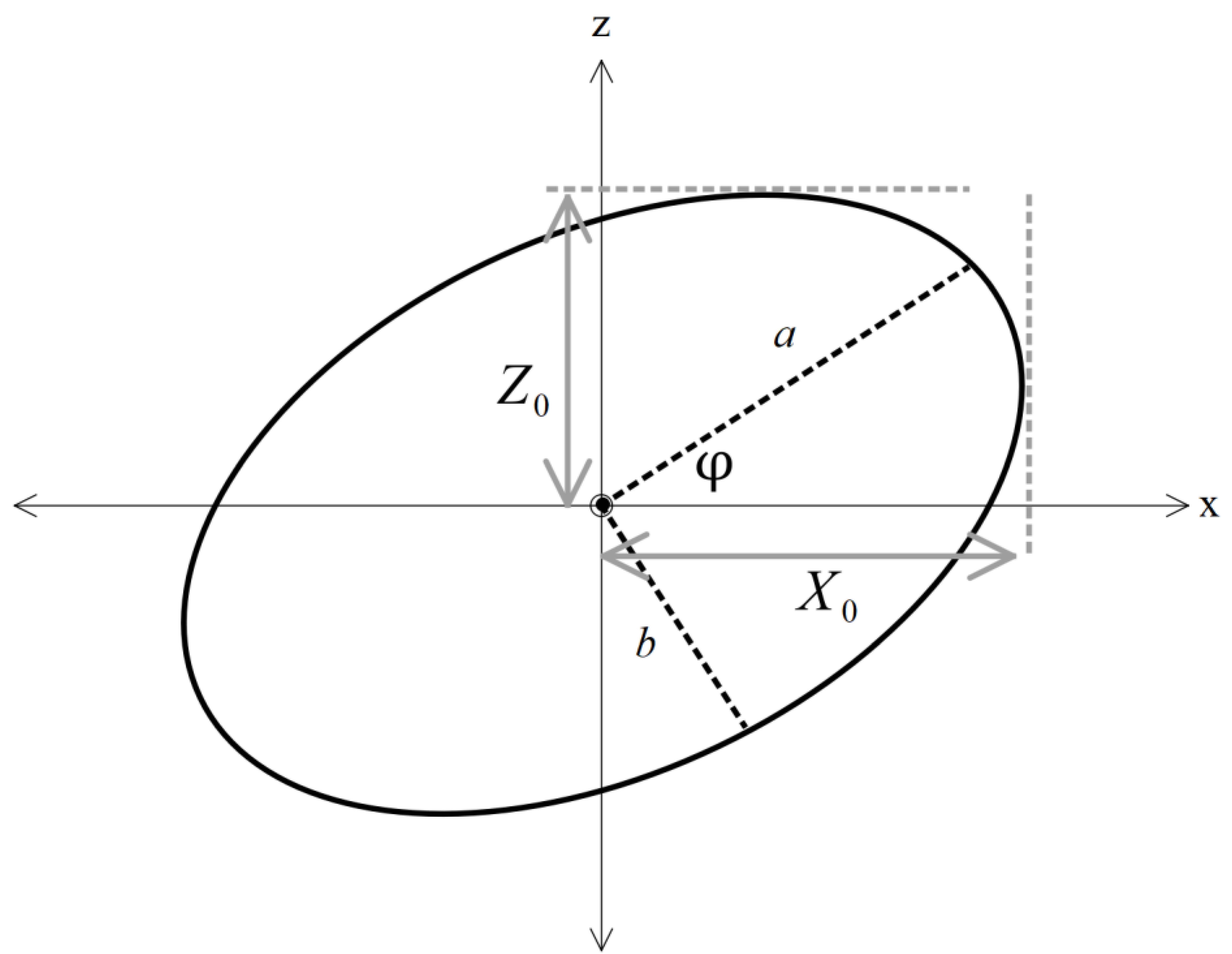}
\caption{motion of the particles at the surface} \label{fig_ellipse}
\end{figure}

\begin{figure}[H]
\centering
\includegraphics[width=12.5 cm]{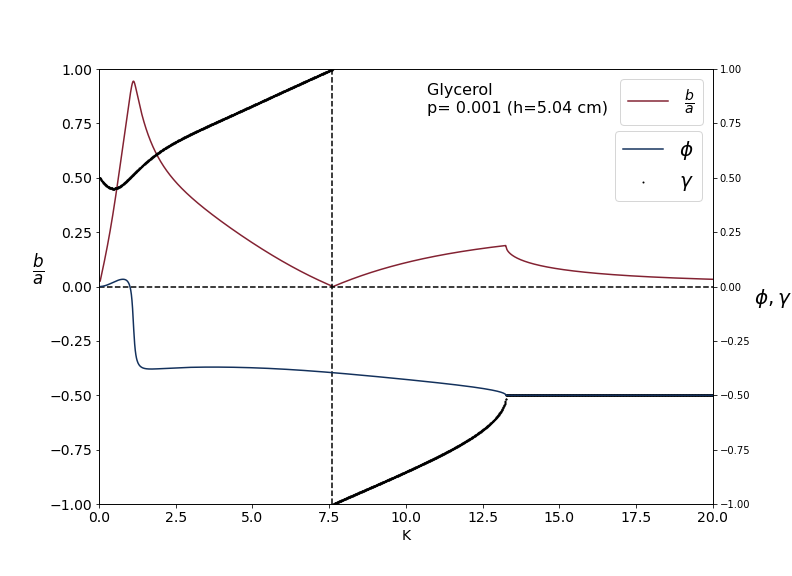}
\caption{The ratio of $\frac{b}{a}$ and the angels $\phi$ and $\gamma$ in terms of $K$ are presented for glycerin at $p=0.001$. } \label{fig_ellipse2}
\end{figure}

In this section, we focus on numerical simulation of particle trajectories associated with wave patterns at the surface of the fluid. Fig. (\ref{fig_ellipse}) illustrates this concept of elliptical motion of fluid particles at the surface, indicating the direction and magnitude of the horizontal and vertical movements. As shown in Fig. (\ref{fig_ellipse2}), the waves move across the surface, By increasing $K$ values fluid particles are moved in a circle, then their trajectory is an counter clockwise elliptical path. the move forwards and backwards along the ellipse's axis   

\begin{align}
& a^{2}=\frac{X_o^{2} \cos ^{2} \varphi-Z_o^{2} \sin ^{2} \varphi}{\cos ^{2} \varphi-\sin ^{2} \varphi} \\
& b^{2}=\frac{Z_o^{2} \cos ^{2} \varphi-X_o^{2} \sin ^{2} \varphi}{\cos ^{2} \varphi-\sin ^{2} \varphi}
\end{align}

Angle of main axis compered to horizontal.
\begin{equation}
    c=\left|\frac{\xi}{X_o}\right|     
\end{equation}

\begin{equation}
    \gamma=\xi-\zeta 
\end{equation}

\begin{align}
    & x=X_o e^{i \zeta} \cdot e^{-i \omega t} \\
    & z=Z_o e^{i \xi} e^{-i \omega t}
\end{align}

\begin{align} \label{ellipse_6}
      X_o e^{i\zeta} =&\frac{i}{\omega} 
 K [ Q(Q^{2}+3 K^{2}) \cdot (\cosh(Q) \cosh(K)-1) \\
 & -K \cdot (K^{2}+3 Q^{2}) \sinh(Q) \sinh(K) ] \nonumber
\end{align}

\begin{align} \label{ellipse_7}
      Z_o e^{i\xi} =&\frac{1}{\omega} 
 K(Q-K^2) \cdot [K \cosh(K) \sinh(Q) \\
 & -Q \cdot \sinh(Q) \sinh(K) ] \nonumber
\end{align}

According to Eqs. (\ref{ellipse_6}) and (\ref{ellipse_7}), when $Q$ is a real, the value of $\gamma$ is $-\frac{\pi}{2}$, but if $Q$ is purely imaginary, then the value of $\gamma$ is $\frac{\pi}{2}$.

\section{Time evolution of surface elevations}

\begin{figure}[H]
\centering
\includegraphics[width=\linewidth]{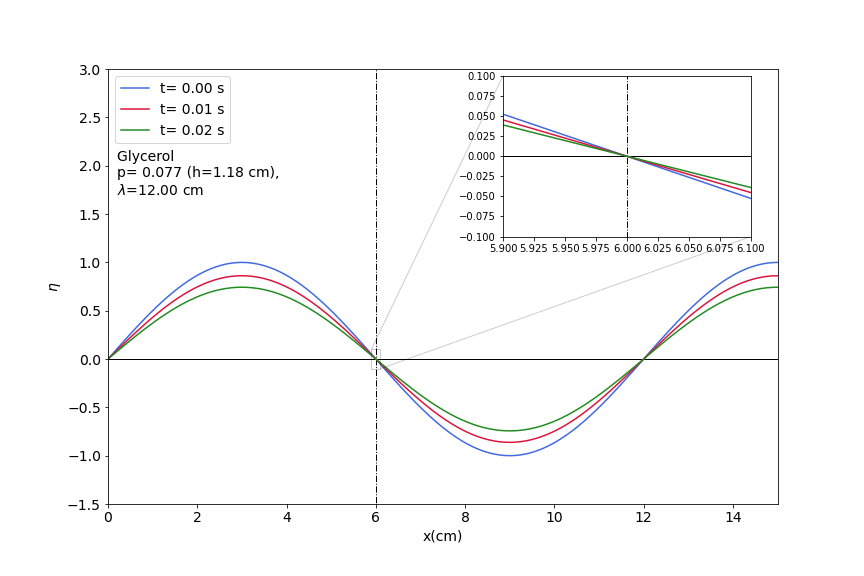}
\caption{Time evolution of surface elevation in glycerin at parameter $p=0.077$ and wave number $K=0.62$.\label{fig_p1}}
\end{figure}

Figs. (\ref{fig_p1})-(\ref{fig_p10}) demonstrate the time evolution of a propagating wave in a fluid, along with the associated dispersion relations. We have chosen glycerin as the fluid medium due to its physical properties, as other fluids may require a thinner layer for observation of the phenomena. Then indeed the critical thickness for glycerin is 1.1 cm and for water is \(3.6\times 10^{-8}\) m and for mercury is \(9.3\times 10^{-10}\) m - this is the reason we have chosen the glycerin for simulation.  

\begin{figure}[H]
\centering
\includegraphics[width=\linewidth]{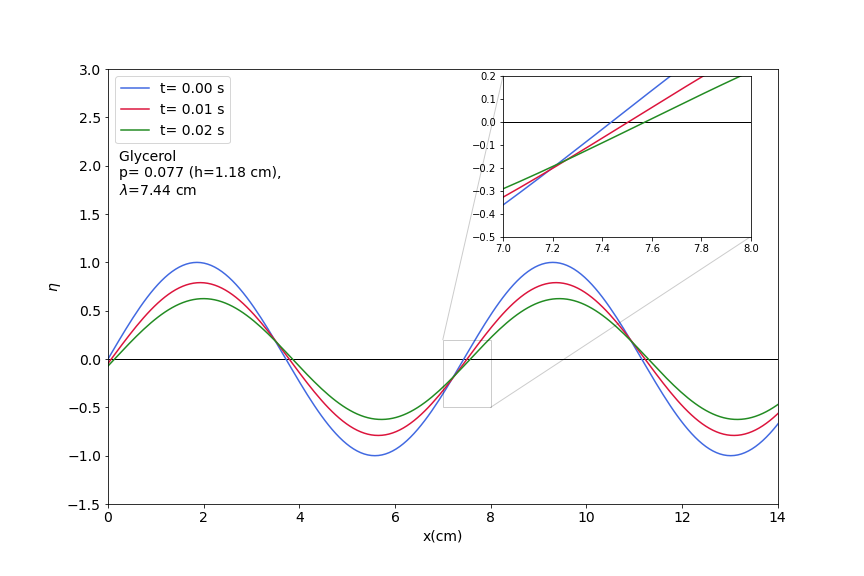}
\caption{Time evolution of surface elevation in glycerin at parameter $p=0.077$ and wave number  $K=1.00$.\label{fig_p2}}
\end{figure}

\begin{figure}[H]
\centering
\includegraphics[width=\linewidth]{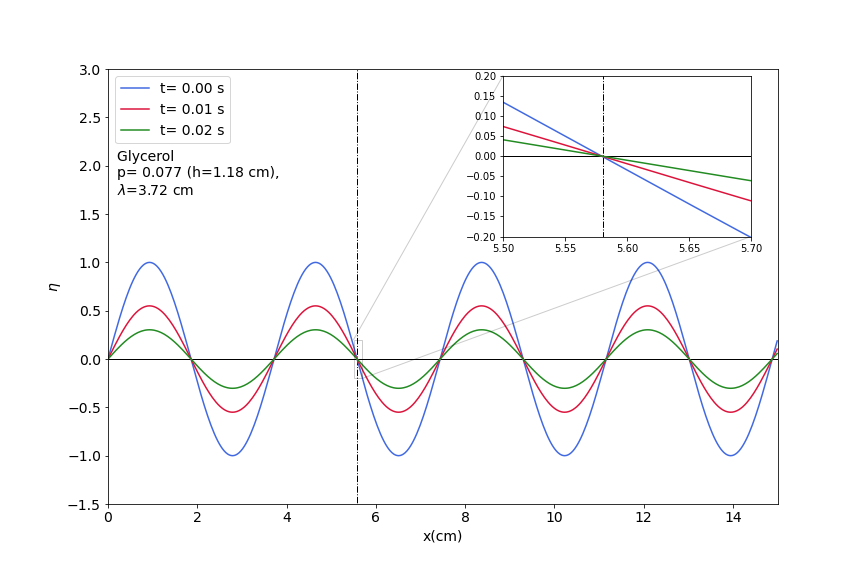}
\caption{Time evolution of surface elevation in glycerin at parameter $p=0.077$ and wave number  $K2.00$.\label{fig_p3}}
\end{figure}

We aim to establish a link between theoretical predictions and experimental observations. Our results show that at $p=0.077$, wave propagation is observed at a specific wave-number $K=1$ (actually in a narrow range around it), whereas no propagation is seen at longer wavelengths, such as $K=0.62$, or at shorter wavelengths, such as $K=2$ (cf. Figs. \ref{fig_p1}-\ref{fig_p3}). At $p=0.001$, however, the range of wavelengths that propagate is broader, although the non-propagating long-wave mode appears at a wavelength of 8 meters in a fluid layer of 5 centimeters, which is challenging to observe (cf. Figs. \ref{fig_p1}-\ref{fig_p10}).

\begin{figure}[H]
\centering
\includegraphics[width=\linewidth]{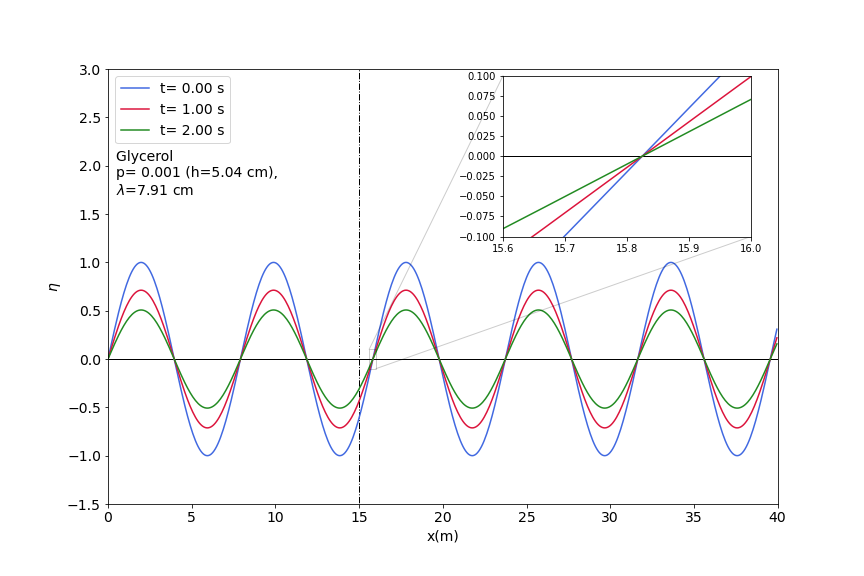}
\caption{Time evolution of surface elevation in glycerin at parameter  $p=0.001$ and wave number $K=0.04$.\label{fig_p5}}
\end{figure}   

\begin{figure}[H]
\centering
\includegraphics[width=\linewidth]{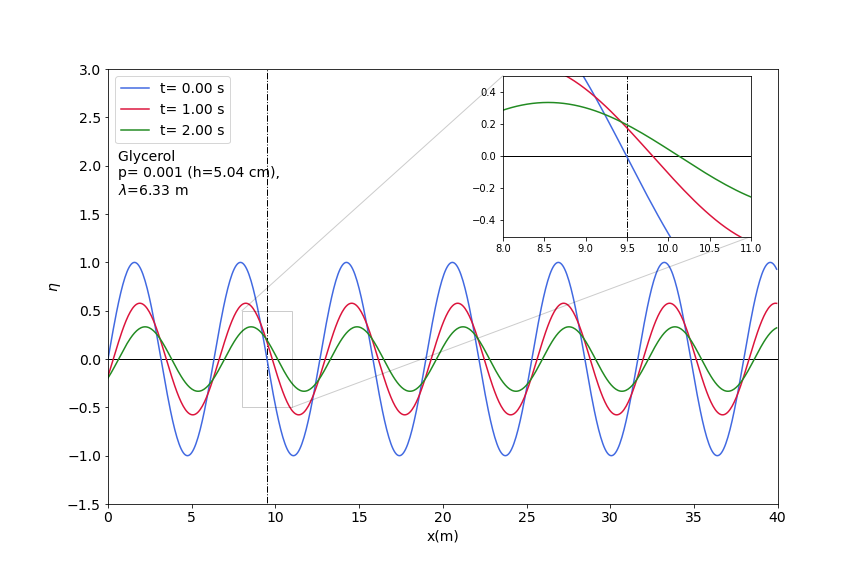}
\caption{Time evolution of surface elevation in glycerin at parameter $p=0.001$ and wave number $K=0.05$.\label{fig_p6}}
\end{figure}   




\begin{figure}[H]
\centering
\includegraphics[width=\linewidth]{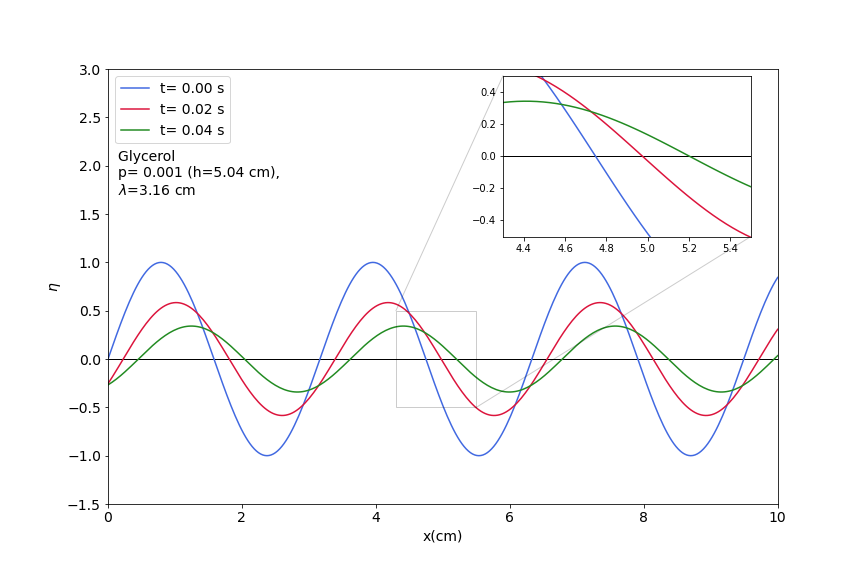}
\caption{Time evolution of surface elevation in glycerin at parameter  $p=0.001$ and wave number  $K=10.0$.\label{fig_p10}}
\end{figure}   

\begin{figure}[H]
\centering
\includegraphics[width=\linewidth]{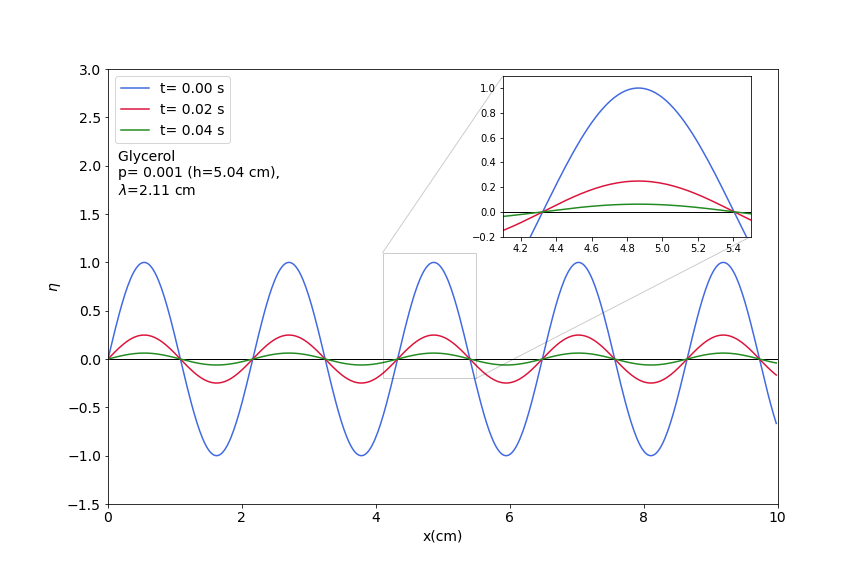}
\caption{Time evolution of surface elevation in glycerin at parameter  $p=0.001$ and wave number  $K=15.0$.\label{fig_p10}}
\end{figure}   

\begin{figure}[H]
\centering
\includegraphics[width=\linewidth]{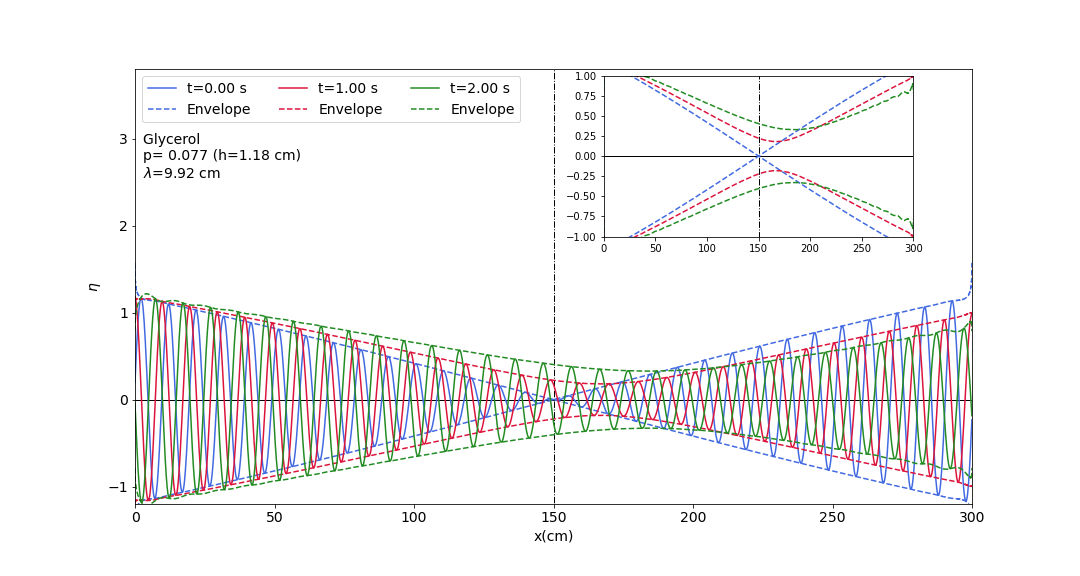}
\caption{Time evolution of surface elevation in glycerin at parameter $p=0.077$ with two nearby wave numbers at $K=0.75$ ($\Delta K=0.01$).\label{fig_p11}}
\end{figure}

\begin{figure}[H]
\centering
\includegraphics[width=\linewidth]{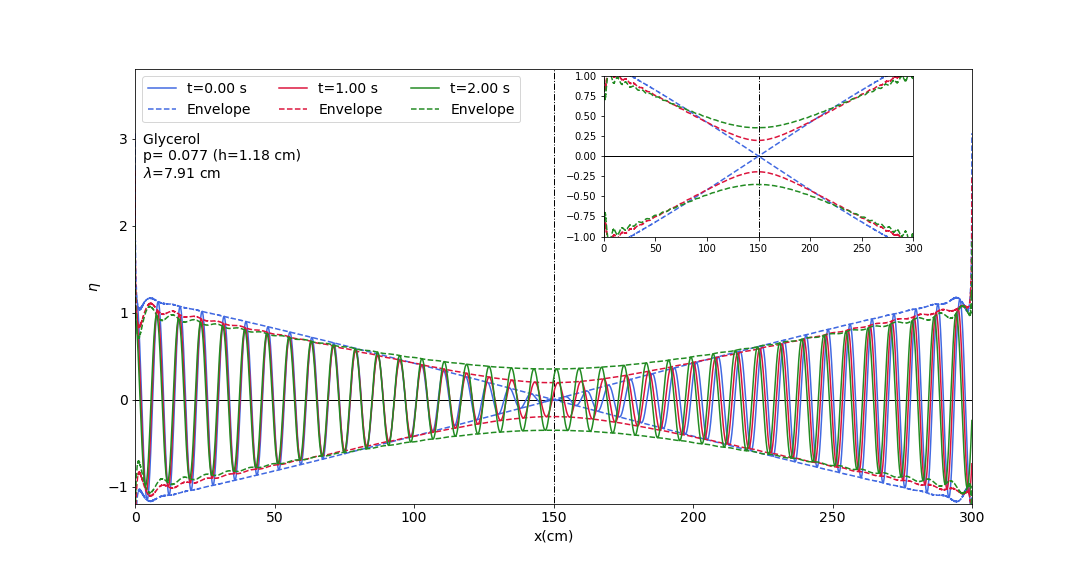}
\caption{Time evolution of surface elevation in glycerin at parameter $p=0.077$ with two nearby wave numbers at  $K=0.94$ ($\Delta K=0.01$).\label{fig_p12}}
\end{figure}

\begin{figure}[H]
\centering
\includegraphics[width=\linewidth]{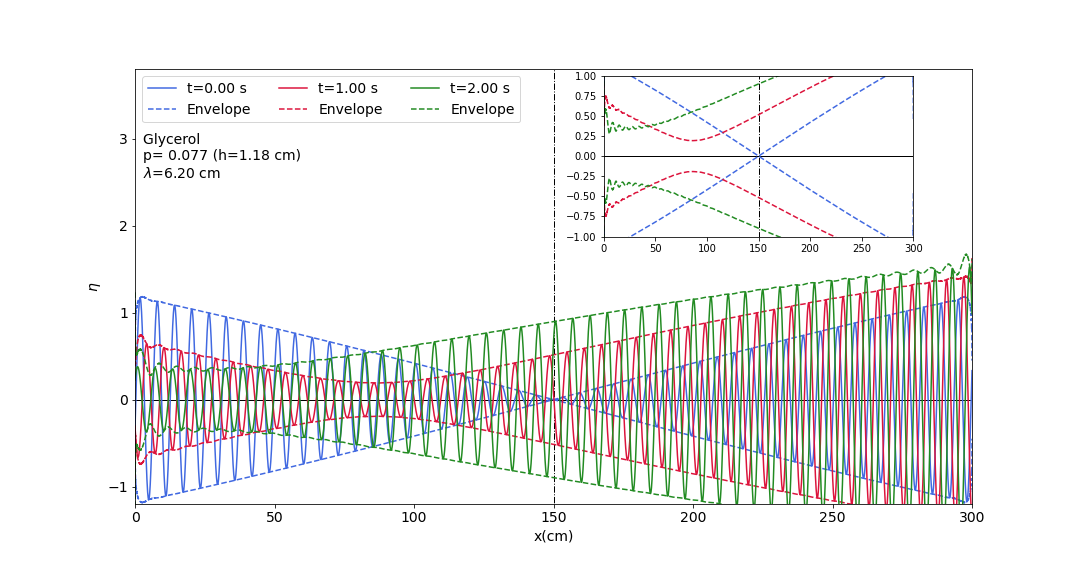}
\caption{Time evolution of surface elevation in glycerin at parameter $p=0.077$ with two nearby wave numbers at  $K=1.20$ ($\Delta K=0.01$).\label{fig_p13}}
\end{figure}

We limit our study to the lowest two branches of the dispersion relation, as shown in Fig. (\ref{fig2}) these branches have the longest lifetime and are the least damped. To illustrate the group velocities in the propagating modes, we superimpose two nearby wavelengths, $\sin(k_1(x-x_o)-\omega_1 t)-\sin(k_2(x-x_o)-\omega_2 t)$, which propagate rightwards. However, direct observation of this phenomenon is unlikely due to strong damping. To visualize the movement of the envelope, we amplify the amplitudes at a rate of $e^{(|\gamma| t)}$, where $\gamma$ is the smaller decay rate. Our results show that positive, zero, and negative group velocities occur as illustrated in Figs. (\ref{fig_p11})-(\ref{fig_p13}), respectively. As a result, the envelope of the wave profile moves to the left while the wave is traveling to the right when the slope of the real part of the omega values is negative (cf. Fig \ref{fig_reim}). Using numerical computations, several snapshots of these important quantities at different times as functions of the wave number are shown.

An arbitrary initial condition means specifying the velocity field at an instant of time everywhere within the fluid layer. In linear approximation one may decompose such an initial condition in terms of modes. Most modes are strongly damped, therefore, leaving them out of the decomposition may not lead to a significant error except initially for a very short time. If we keep only the lowest two branches, it is possible to formulate the initial value problem in terms of the surface profile and its time derivative. Note that in the range of wave numbers where propagation is possible, the two branches differ only in the sign of the real part of the frequency, allowing a description of both direction of propagation. Explicitly, we may formulate the initial value problem in wave number space (Fourier space) as follows. At a given wave number we have two modes, therefore the decomposition is

\begin{equation}
c_1(k) e^{-i \Omega^{(1)} t}+ c_2(k) e^{-i \Omega^{(2)} t}
\end{equation}

for surface elevation at time t.
If the initial surface profile is $\eta(k,t=0)$, then
\begin{equation}
    \eta(k,t=0)=c_1+c_2
\end{equation}

should hold. Similarly, given the initial vertical velocity profile $\eta_t(k,t=0)$ we have

\begin{equation}
    \eta_t(k,t=0)=-i \Omega^{(1)} c_1-i \Omega^{(2)} c_2
\end{equation}
From this we get for the coefficients $c_1$ an $c_2$
\begin{align}
  c_1&=\frac{i \eta_t-\Omega^{(2)} \eta}{\Omega^{(1)}-\Omega^{(1)}}  \\
  c_2&=\frac{i \eta_t-\Omega^{(1)} \eta}{\Omega^{(2)}-\Omega^{(1)}}
\end{align}

This allows one to solve the initial value problem within the limits of the approximation sketched above. On the other hand, such an approximation is completely equivalent with a second order differential equation for the surface elevation

\begin{equation}
    \frac{\partial^2 \eta}{\partial t^2}+i (\Omega^{\left ( 1 \right )}\left ( k \right )+\Omega^{\left ( 2 \right )}\left ( k \right )) \frac{\partial \eta}{\partial t}-\Omega^{\left ( 1 \right )}\left ( k \right ) \Omega^{\left ( 2 \right )}\left ( k \right ) \eta =0
\end{equation}

\begin{figure}[H]
\centering
\includegraphics[width=12.5 cm]{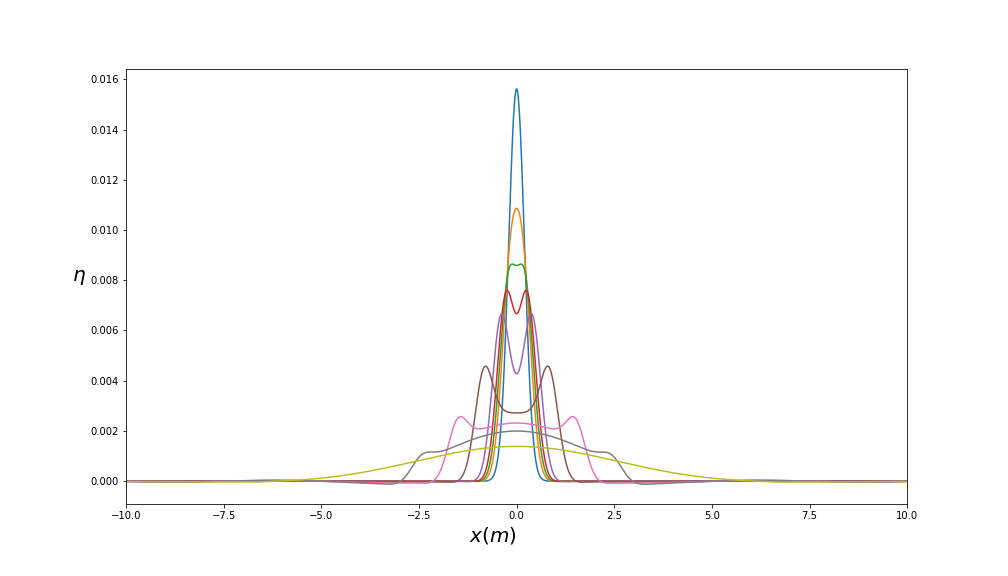}
\caption{Time evolution of an initial Gaussian wave for glycerin at $p=0.001 (h=5.04 cm)$ presented. Time values are t = 0., 0.3, 0.4, 0.5, 0.7, 1.4, 2.5, 4., and 10. s. A slim initial Gaussian wave consist of propagating modes (blue line) spread outwards and gradually diminish and left behind a broader Gaussian shape profile due to non-propagating modes at long wavelengths (yellow line). }\label{fig_evolution}
\end{figure}

Fig. (\ref{fig_evolution}) shows fascinating aspect of non-propagating modes at long wavelengths. As demonstrated, narrow initial Gaussian wave encompassing propagating modes (blue line) transformed into to two peaks radiate off symmetrically due to propagating modes. Since there is no initial velocity, peaks are in a same weight. Gradually, the Gaussian wave gives way to a much broader Gaussian shape profile, composed of non-propagating modes (yellow line).

\section{Conclusions}

Linear viscous capillary-gravity waves were studied in a channel of constant depth, without restricting the parameters. We explored all the modes numerically. Modes were labelled by horizontal wave number \(K\), a continuous parameter, and vertical wave number \(Q\), a discrete, complex quantity. We found that there were always infinitely many nonpropagating modes. Propagation can only occur in the two modes with smallest decay. In a sufficiently thin layer no propagation occurs at all. When increasing layer thickness, a bifurcation occurs which shows up in the plot of imaginary parts of frequencies like a collision of the lowest lying branches of modes. After that collision, at increasing wave numbers one can observe a merging and a subsequent split of these branches. In the wave number range of the merged section frequencies  have nonzero real part (propagation). We stress that even at those depth where propagation becomes possible, propagation at very low and very high wave numbers are still prohibited. The latter is already known \cite{ref06_armaroli2018viscous}, but the former seems to be our finding. We also determined the minimal layer thickness necessary to wave propagation. Further, we studied surface motion. Assuming a monochromatic wave propagating to the positive \(x\) direction we found that a surface particle in a viscous fluid could rotate both clockwise or counterclockwise, depending on the wave number. We also demonstrated the propagation or non-propagation of waves in a few cases. In order to illustrate that both positive, zero or even negative group velocities can occur, the beat of two nearby wave numbers was displayed at a few consecutive time instants. Finally, with the assumption that the effect of fast decaying high lying branches was negligible, we kept the lowest two modes and formulated the solution of the initial value problem of surface motion in wave number space. As an application, time evolution of a narrow initial Gaussian surface elevation with zero velocity was studied and a radiation of propagating modes in the form of two oppositely travelling bumps was observed. A slowly decaying wide Gaussian was left behind, consisting of large wavelength nonpropagating modes.     

\section{Acknouledgement}
This research was supported by the Ministry of Culture and Innovation and the National Research. A.Gh. greatly acknowledges the support from Stipendium Hungaricum.

\clearpage

\bibliographystyle{unsrt}
\bibliography{main}

\end{document}